\documentclass[11 pt]{article}%
\usepackage{bbm}
\usepackage[longnamesfirst]{natbib}
\usepackage{amssymb}
\usepackage{amsmath}
\usepackage{amsthm}
\usepackage{setspace}
\usepackage[colorlinks]{hyperref}
\usepackage{graphicx}
\usepackage{caption}
\usepackage{subcaption}
\usepackage{enumitem}
\usepackage[margin=1in]{geometry}
\usepackage{verbatim}
\usepackage{sectsty}
\usepackage[svgnames]{xcolor}
\usepackage{subfiles}
\usepackage{xargs}
\usepackage{titlesec}
\usepackage[colorinlistoftodos,textsize=tiny]{todonotes}
\usepackage[capitalize,nameinlink]{cleveref}
\usepackage[normalem]{ulem}
\usepackage[font={small,it},justification=justified]{caption}
\usepackage{xcolor,colortbl}
\usepackage{pgf}
\usepackage{tikz}
\usepackage{amsfonts}%
\providecommand{\U}[1]{\protect\rule{.1in}{.1in}}
\setlist{noitemsep,parsep=6pt,partopsep=0pt,topsep=0pt}
\hypersetup{
pdftitle={Q-learning with biased policy rules},
pdfauthor={Olivier Compte},
citecolor=DarkBlue,
bookmarksnumbered=true,
urlcolor=Indigo,linkcolor=DarkBlue
}
\theoremstyle{remark}

\theoremstyle{plain}

\renewcommand{\epsilon}{\varepsilon}
\usetikzlibrary{patterns, decorations.pathreplacing, arrows.meta,calc}
\let \savenumberline \numberline
\def \numberline#1{\savenumberline{#1.}}
\makeatletter
\renewcommand\@seccntformat[1]{\csname the#1\endcsname.{\hskip.7em\relax}}
\makeatother


\let\oldfootnote\footnote
\renewcommand\footnote[1]{\oldfootnote{\hspace{.5mm}#1}}
\titlespacing\section{0pt}{10pt plus 2pt minus 2pt}{4pt plus 2pt minus 2pt}
\titlespacing\subsection{0pt}{6pt plus 2pt minus 2pt}{2pt plus 2pt minus 2pt}
\titlespacing\subsubsection{0pt}{6pt plus 2pt minus 2pt}{0pt plus 2pt minus 2pt}
\titlespacing{\paragraph}{  0pt}{  0.5\baselineskip}{  1em}
\newcommand{\appendixref}[1]{\hyperref[#1]{Appendix \ref{#1}}}
\definecolor{dark-red}{rgb}{0.4,0.15,0.15}
\definecolor{dark-blue}{rgb}{0.15,0.15,0.75}
\definecolor{medium-blue}{rgb}{0,0,0.5}
\usetikzlibrary{shapes,arrows,automata,fit}
\tikzstyle{info}=[circle,thick,draw=black,fill=black!25,minimum size=4mm]
\tikzstyle{uninfo}=[circle,thick,draw=black,fill=white,minimum size=4mm]
\tikzstyle{inforecog}=[circle,line width=1mm,draw=black!50,fill=black!25,minimum size=4mm]
\tikzstyle{uninforecog}=[circle,line width=1mm,draw=black!50,fill=white,minimum size=4mm]
\tikzstyle{traded}=[draw, line width=1mm]
\tikzstyle{recog}=[draw=black!50, line width=1mm]
\sectionfont{\color{DarkRed}}
\subsectionfont{\color{DarkRed}}
\subsubsectionfont{\color{DarkRed}}
\newcommandx{\nageeb}[2][1=]{\todo[linecolor=blue,backgroundcolor=blue!25,bordercolor=blue,#1]{#2}}
\newcommandx{\andreas}[2][1=]{\todo[linecolor=black,backgroundcolor=black!25,bordercolor=black,#1]{#2}}
\newcommandx{\navin}[2][1=]{\todo[linecolor=red,backgroundcolor=red!25,bordercolor=red,#1]{#2}}
\begin{document}
\begin{titlepage}
\title{Learned Collusion}
\date{May 27$^{th}$ 2025}
\author{Olivier Compte\thanks{Affiliation: \textit{Paris School of Economics}, 48
Boulevard Jourdan, 75014 Paris and \textit{Ecole des Ponts Paris Tech},
\href{mailto:olivier.compte@gmail.com}{\color{dark-blue}olivier.compte@gmail.com}.  I thank Philippe Jehiel for helpful discussions. This version supercedes previous versions entitled ``Q-based equilibria'' and ``Q-learning with biased policy rules".}}
\maketitle
\begin{abstract}
\noindent
Q-learning can be described as an all-purpose automaton that  provides estimates (Q-values) of the continuation values associated with each available action and follows the naive policy of almost always choosing the action with highest Q-value. We consider a family of automata based on Q-values, whose policy may systematically favor some actions over others, for example through a bias that favors cooperation. We look for stable equilibrium biases, easily learned under converging  logit/best-response dynamics over biases, not requiring any tacit agreement. These biases strongly foster collusion or cooperation across a rich array of payoff and monitoring structures, independently of initial Q-values.
\noindent
Keywords: Artificial Intelligence, Learning, Bounded Rationality, Cooperation
\noindent
JEL Classification Codes: C63, C73, D43, D83, D9, L51
\end{abstract}
\thispagestyle{empty}
\end{titlepage}

\setstretch{1.1}

\clearpage
\clearpage
\setcounter{page}{1}

\onehalfspacing

\section{Introduction}

Long-term relationships have been extensively studied in economics, aimed at
understanding how repeated interactions can foster cooperation. In a large
strand of the economic literature, the methodology has been mostly
constructive: given a stage game characterized by its payoff and information
structures, the repeated game analysis often consists in finding, within the
huge set of feasible strategies of the repeated game, strategies that work
well, i.e., \textit{design} strategies that, when played, support a given
level of cooperation, say, and also ensure that each player has incentives to
play their part all along. This design perspective has been at the heart of
the Folk Theorem results in the literature.\footnote{See \citet{fudenberg86},
\citet{abreu90}, \citet{fudenberg94}, \citet{sugaya22}, or \citet{mailath06}
for an overview of the literature.}

In another strand of the literature, inspired by computer science, the
methodology
departs significantly from this design perspective. It starts from an
\textit{exogenously fixed} all-purpose automaton defined independently of the
game considered (e.g., a reinforcement learning algorithm -- Q-learning), and
then it examines, across a variety of games, the consequence for mutual play
when each player uses such an automaton (\citet{sandholm96}). In this vein,
the recent work of \citet{calvano20} and \citet{banchio22a}) finds that
Q-learning automata can be good strategies for enabling collusion or
cooperation, at least for some payoff structures (and favorable initial
conditions).

Regarding policy implications, the latter contributions are important because,
in contrast to the classic repeated game approach, which designs strategies
tuned to the specific game considered, Q-learning allows players to sustain
significant and persistent cooperation without them having any prior knowledge
of the game structure. Still, there are caveats. The agents are assumed to
have no discretion over the strategy/automaton used (even though, as we shall
see, Q-learning can be far from being optimal). Cooperation is only obtained
for some payoff structures where gains from deviating are moderate and it
requires favorable initial conditions (high enough initial Q-values). This may
drastically reduce the scope for collusion under Q-learning if payoffs are
subject to partially persistent shocks that drive the dynamics to an absorbing
non-cooperative phase.

Our objective is to partially address these issues, keeping the spirit of the
algorithmic literature: focusing on a family of all-purpose automata (rather
than a single one), we allow for discretion over the choice of automata within
the family, we check whether this is conducive to persistent cooperation
\textbf{independently of initial conditions} on Q-values, and we do that for a
large array of payoff and monitoring structures. Furthermore we check that the
automaton eventually selected is \textbf{easily learned} under best or
logit-response dynamics within the family of automata considered.

Specifically, our automata use the same $Q$-values as those used by the
Q-learning automaton. But rather than assuming that the agent's decision rule
consists in selecting the action with the currently highest $Q$-value, we
allow for
a strategic dimension taking the form of a (one-dimensional) systematic bias
that favors some alternatives over others. We call this $Qb-$\textit{learning}%
, where $b$ could stand for \textquotedblleft biased" or \textquotedblleft
based". Each bias chosen by player $i$ defines a personal automaton. Each bias
profile induces a long-run payoff for each player. Assuming that each player
sets her own bias non-cooperatively, we obtain a game over biases. We are
interested in equilibrium bias profiles, refer to them as $Qb-$%
\textit{equilibria} and examine their evolutionary
stability.\footnote{Following \citet{compte23}, we do that by computing the
limit quantal response equilibria (limit QRE), obtained by finding the highest
precision for which the logit-response dynamics remains stable: starting from
low precision, and gradually increasing it, the procedure allows us to make a
single equilibrium prediction.}
One lesson we draw is that, compared to \citet{banchio22a}, this option to
bias the decision rule strongly broadens the scope for cooperation or
collusion: in the equilibria selected, each player adopts a bias that induces
persistent cooperation independently of initial conditions. Furthermore, this
finding extends to cases where players' rewards are stochastic and independent
(conditional on the action profile played) as is the case in the literature on
repeated games with imperfect private monitoring. Another lesson is that these
collusive biases are easily learned individually, and not requiring any tacit
agreement: there are strong evolutionary pressures towards them.

Our approach is in the spirit of \citet{compte18}, who advocates the use of
strategy restrictions to model moderately sophisticated players. Here the
restriction comes from the assumption that behavior is $Q$-based and the only
strategic dimension allowed is the constant bias a player uses to determine
behavior at any date, given $Q$-values. Said differently, an automaton defines
a potentially complex relationship between past observations and behavior. The
moderate sophistication means that within the universe of possible algorithms,
the set of algorithms compared is limited (i.e., restricted to those that are
based on $Q$-values, up to a one-dimensional bias). At some broad level, our
approach has the flavor of evolutionary game theory where only a limited
number of strategies compete (as in \citet{axelrod84}): the set of possibly
biases defines a family of possible personal automata (i.e., a family of
possible strategies for the entire game), and, within an ecology of such
strategies, one expects that evolution eventually selects an equilibrium bias.

Formally, in its memoryless version, a $Q$-learning algorithm is an automaton
that uses past experience to calculate a $Q$-value $Q_{i}^{t}(a_{i})$ for each
action $a_{i}$ that player $i$ may play at stage $t$ of the game.\footnote{A
more sophisticated algorithm could compute the $Q$-values $Q_{i,h_{i}}%
^{t}(a_{i})$, \textit{conditional on the recent history }$h_{i}$%
\textit{\ observed by }$i$. Althouth we restrict attention to memoryless
algorithms, our definitions could apply to these more general versions of
$Q$-learning.} Calling $r_{i}^{t}$ the reward obtained at $t$, $Q-$values are
updated according to:\footnote{We use the normalization $(1-\delta
)r+\delta\max Q$ in order to normalize $Q$-values to stage-game values, as is
standard in the repeated game literature. Note that we assume no updating of
$Q_{i}(a)$ when $a$ is not played. This is called asynchronous Q-learning:
players do not attempt to draw inferences about the payoffs they would have
obtained had they played differently nor use these inferences to update $Q$
values.}
\begin{equation}
Q_{i}^{t+1}(a_{i})=(1-\alpha)Q_{i}^{t}(a_{i})+\alpha((1-\delta)r_{i}%
^{t}+\delta\max_{x_{i}}Q_{i}^{t}(x_{i})) \label{Q}%
\end{equation}
when $a_{i}$ is played, and
\[
Q_{i}^{t+1}(a_{i})=Q_{i}^{t}(a_{i})\text{ otherwise.\qquad\qquad\qquad
\qquad\qquad}%
\]
As time goes by, these $Q$-values are updated many times for all $a_{i}$
because the agent is assumed to experiment with positive probability in every
period: in its $\varepsilon-$greedy version, experimentation occurs with
probability $\varepsilon$ in every period and any feasible action is then
selected with same probability.\smallskip

The $Q$-value $Q_{i}^{t}(a_{i})$ can be interpreted as the \textit{subjective}
evaluation of choosing $a_{i}$ at $t$, and it is often thought of being a
proxy for the continuation value associated with playing $a_{i}$ at $t$. Given
$Q$-values, the \textquotedblleft naive\textquotedblright$Q$-learner then
chooses (unless she experiments) an action $a_{i}^{t}$ which maximizes the
$Q-$value. Instead, we allow the agent to select an action that maximizes a
biased criterion:%
\[
a_{i}^{t}\in\arg\max_{a}\ Q_{i}^{t}(a_{i})+b_{i}G_{i}(a_{i}),
\]
where $b_{i}$ is a systematic one-dimensional bias and $G_{i}(a_{i})$ is an
exogenous distortion that may reflect the agent's broad understanding of the
structure of the game, or some natural ordering on strategies possibly based
on broad characteristics of the payoff structure of the game, we shall come
back to that in Section~\ref{sec_duopoly}.\footnote{In games with only two
actions $a_{i}\in\{0,1\}$, the shape of the distorsion plays no role, one can
set $G_{i}(a_{i})=a_{i}$.} Each bias $b_{i}$ thus defines a particular
$Q$-based automaton, and the automaton which adopts the naive policy of
setting $b_{i}=0$ corresponds to standard $Q$-learning.

Our motivation for introducing $Q$-based automata that may be biased (or
biased $Q-$learning), is that in strategic environments (but also in
non-strategic environment where there is a hidden state variable), $Q$-values
may provide a biased estimate of the long-run benefits of choosing a
particular action, so the variable $b_{i}$ can actually be seen as an
instrument, which, if appropriately set, may allow a player to
\textit{de-bias} $Q$-values, to some extent, and improve her welfare.

Said differently, the standard Q-learning automaton (which sets $b_{i}=0$) may
deliver sub-optimal gains. It is thus reasonable to assume that players would
search for (and find) a superior automaton. In practice, this is typically
achieved by firms through so-called AB tests between the current automaton and
a modified one, checking whether using the latter one would not raise revenues.

We will not be modelling the search for the gain-maximizing automaton, but
assume, as is standard in equilibrium analysis, that players manage to find
which automaton (i.e., which bias) maximizes her long-run
payoff.\footnote{Practically, we will approximate ex ante long-run payoffs by
running simulations over \textit{a fixed horizon}, starting from an initial
condition which we will choose to be \textit{unfavorable} to cooperation.}
Note that the justification for optimization over possible automata that we
invoke here is not based on some knowledge of the behavior of others or any
complex computation, but rather the result of \textit{learning from
experience} that one automaton generates more revenues than another one, with
each agent trying to improve upon the choice of their automaton (in a limited
way, i.e., within a limited family of automata). Our approach is thus
consistent with the standard evolutionary justification for Nash equilibrium.
Compared to standard equilibrium analysis, the only qualification is that
optimization is done over a subset of automata.
\medskip

\textbf{Biased Q-values and Q-traps. }How can Q-valued be biased? Consider a
\textit{single} agent facing an environment where payoffs are
\textit{state-contingent} and where the evolution of the state is hidden and
partially governed by the actions played.
Then $Q-$values do not necessarily give adequate guidance on which actions to
play: the agent may be stuck in what we call a\textit{\ }$Q$\textit{-trap},
where $Q$-values suggest using actions that keep the dynamics in low-rewarding
states, with occasional exploration not permitting exit from $Q$-traps. Higher
experimentation levels $\varepsilon$ combined with higher speed of adjustment
$\alpha$ may permit exits, but these higher $\varepsilon$ and $\alpha$ may be
conducive to lower welfare and faster return to the $Q-$trap.
Section~\ref{sectionQtraps} provides a simple class of examples of this
kind.\footnote{See \citet{singh94} or more recently \citet{barfuss22} for
other examples where Q-learning is suboptimal.}

In games such as the repeated prisoner's dilemma or more generally repeated
games of price or quantity competition, a $Q$-trap arises when all players
start choosing persistently a non-cooperative action (defection, low price or
high quantity): cooperation may be sustained for some (possibly long) time
when initial conditions are favorable (as in \citet{banchio22a}), but if
initial conditions are not favorable (or for payoff structure conducive to
Q-traps), learning to recoordinate on cooperation may be hard, with
non-cooperative phases becoming the preponderant ones.

In such circumstances, we obtain $Qb-$equilibria that are more cooperative
than the outcome generated by naive $Q-$learning.

\textbf{Why biased Q-learning helps}. First note that the choice of bias is
strategic in our framework, so there is no guarantee that $Qb-$equilibria lead
to more cooperation. As a matter of fact, the opposite could be true, as
biasing one's policy towards \textit{less} cooperation could be profitable: if
this more severe attitude does not undermine too much the sustainability of
cooperation, it could in principle lead to more gains.\smallskip

Intuitively, Qb-learning helps for the following reason. Under naive
$Q-$learning, a typical path of play consists of an alternation between
\textit{high}$-Q$\textit{\ and low}$-Q$\textit{\ phases.} Low $Q-$phases are
traps where players mostly defects. High-$Q$ phases consists of (relatively
frequent) alternations between \textit{jointly cooperative phases} (which
boost $Q$-values) and \textit{disorganized phases} that mostly consist of
$CD$'s and $DD$'s (which depresses the $Q$-values of both players). The role
of these disorganized phases is to realign $Q$-value differences across
players, to facilitate simultaneous re-coordination on cooperation. Sometimes
these disorganized phases are too long and players end up in a low-Q phase,
from which exit takes time and requires simultaneous experimentations.

Biased-Q learning is privately (and socially) helpful because
it reduces the chance of falling in a low-Q phase and shortens these low-Q
phases if they arise.

Note however that while biased Q-learning helps even individually, there is an
equilibrium constraint: biases cannot be too high. If one's opponent is too
strongly biased, then using the standard (unbiased) $Q-$learning automaton
will be a better strategy: through occasional experimentation, this unbiased
automaton will eventually find that defection is a better alternative (hence
take advantage of her highly biased opponent).

\medskip

\textbf{Monitoring technology.} There is a long-tradition in economics of
examining repeated games according to their monitoring technology,
distinguishing between perfect (observable action profiles), imperfect public
(observable public signals correlated with actions), or\ imperfect private
(private signals correlated with actions -- and typically independent
conditional on the actions) monitoring.
This distinction has been useful in structuring the study of these games as
the monitoring technology shapes the strategies available and the easiness
with which dynamic programming techniques can be used to solve them
(\citet{abreu90}): both perfect and public monitoring allow the use of public
strategies where computing continuation values after any history can be done
perfectly (by the analyst).

With $Q$-learning, this distinction between public and private monitoring
seems irrelevant. $Q$-values provide a\textit{\ private} summary statistic of
one's past payoffs, so $Q$-based strategies are not public -- they use private
information. Furthermore, it would seem that the nature of monitoring should
not matter much: whether monitoring is perfect, imperfect, public or private,
some averaging of past payoffs is going on, so this should not affect much
behavior: imperfect signals (e.g., stochastic payoff realizations) merely add
some randomness into $Q-$values.

What our analysis reveals however is that this randomness affects the
evolution and co-evolution of $Q-$values across players in ways that affect
the chance of falling and staying in a $Q-$trap. In particular, shocks on
payoffs deteriorate the ability of players to sustain cooperation durably, but
most of all, \textit{independent }shocks deteriorate the ability to exit from
traps (when players use the naive Q-learning automaton). Exit from traps
requires coordinated moves, and these are more difficult to generate when
shocks are independent.

Given this difficulty, the option to bias $Q-$learning has great potential for
helping players sustain cooperation, and this is what
Section~\ref{sec_stochastic} confirms. We obtain equilibrium biases that
sustain an equivalently high level of cooperation whether payoffs are
deterministic or not, and whether shocks are correlated or independent,
independently of initial conditions on $Q$-values.


The paper is organized as follows. Section~\ref{sec_lit} discusses the related
literature. Section~\ref{sectionQtraps} presents a dynamic decision problem in
which $Q-$learning performs poorly. Section~\ref{section3} considers the
repeated prisoner's dilemma, explaining first the dynamics of play and the
various phase alternations (cooperative, disorganized and defective), and then
the role of biases in mitigating the occurrence of defective phases.
Section~\ref{sec_stochastic} introduces stochastic payoffs, while
Section~\ref{sec_duopoly} applies our analysis to the oligopoly setup studied
in \citet{calvano20}. Section~\ref{sec_conclusion} concludes.

\section{Related literature\label{sec_lit}}

\textbf{Some methodological comments.} We discuss further how our work stands
compared to the two strands of the literature mentioned earlier: the classic
one aiming for Folk Theorems, and the algorithmic one.

A central question addressed by the classic repeated game literature is
whether players have incentives to adopt history-dependent behavior that
fosters cooperation, e.g., whether punishments are credible: if a player
believes that her opponent will cooperate in the future, why would he penalize
her for past wrongdoings? This credibility issue is addressed through the
design of well-coordinated continuation strategies.

For example, when public signals correlated with actions are available, past
public signals can be used as a coordinating device to modify the course of
play of all players simultaneously (\citet{abreu90} and \citet{fudenberg94}).
Then, behavior depends on the history of signals because players
\textquotedblleft agree" to coordinate play in this way. In particular,
players do not use past signals to learn about other's past actions, because
these past signals (and not past actions) fully determine other's future
behavior. These coordination possibilities facilitate the design of
equilibrium strategies, allowing the analyst to focus on whether a given value
vector can be enforced with appropriate continuations values (which should
themselves be enforceable).\footnote{When signals are private, coordination
possibilities are more limited. Still, following the belief-free approach
(\citet{piccione02} and \citet{ely02}), and to the extent that players share a
common clock, one can design strategy profiles that \textquotedblleft restart"
at predetermined dates, with play conditioned on recent private signals only
(\citet{sugaya22}). At these predetermined dates, continuations strategies are
designed so that each player is indifferent between various paths of play.
Given these indifferences, there is scope for conditioning continuation play
on recent past signals in an incentive compatible way.}


While public events may certainly play a role in practice in shaping mutual
beliefs, this literature does not investigate a possibly more obvious motive
for history dependence, the fact that it could be a simple by-product of each
individual independently and optimally learning from past data in an uncertain
environment: if there is some uncertainty about what others do, and some
persistence in what others do, then it may be worth using past data to better
adapt one's behavior to others'.
For example, if there are exogenous and persistent shocks on whether
cooperation is worthy (as in \citeauthor{compte18}~(\citeyear{compte18},
\href{http://www.parisschoolofeconomics.com/compte-olivier/Chapter14Cooperation.pdf}{\color{dark-blue}Ch.
14}), optimal processing of information should naturally give rise to
history-dependent behavior. Within such stochastic environments however, the
characterization of fully optimal strategies (i.e., Bayesian learning) is
particularly difficult.

The algorithmic literature takes a quite different route, assuming history
dependence from the outset: it defines a specific (learning) algorithm for
treating past observations and playing each stage game, that is, a specific
strategy for the entire game.
For Q-learning, it corresponds to a \textit{non-Bayesian learning rule} which
aims to pick the optimal action at any stage, using Q-values to subjectively
estimate the long-run consequences of each action at the current stage. One
motivation for using such a strategy is that it performs well in simple
decision theoretic environments. But there is of course no guarantee that it
does so in the strategic environments that we consider. The algorithm is
fixed, its optimality within the set of possible strategies/algorithms is not
examined: the analyst does not check whether the agent has indeed
\textit{objective} incentives to cooperate at all dates where $Q^{t}%
(C)>Q^{t}(D)$ and defect otherwise.

In this paper, we check incentives, but only partially, within a simple class
of repeated-game strategies where the decision rule at each date is
systematically distorted by a one-dimensional bias.\footnote{In the Appendix,
we briefly examine the effect of adding more discretion, allowing each agent
to modify their own speed of adjustment $\alpha$.} In other words, given the
stochastic environment that a player faces, there is optimal learning, but
only within a quite limited set of learning rules. This is \textit{boundedly
optimal learning}, short of Bayesian learning.

Summarizing this discussion, our work contributes to the classic repeated game
literature, showing that whether monitoring is perfect, imperfect public or
private, cooperation can arise without relying on a sophisticated coordination
of continuation strategies, but simply out of each player independently
attempting to learn optimally how to behave in a stochastic environment, with
the caveat that the search for an optimal learning rule is constrained to a
limited set of Q-based rules.
We also contribute to the Q-learning literature in that we do not take for
granted that players would stick to an exogenously given (and possibly
suboptimal) non-Bayesian learning rule, but investigate the consequence of
allowing for some discretion over the learning rules.

\textbf{Contribution to Q-learning.} We now turn to our specific contribution
to the algorithmic literature and the rapidly growing body of work that
examines how effective algorithmic strategies based on reinforcement learning
can be at supporting collusion or cooperation (\citet{calvano20,calvano21},
\citet{klein21}, \citet{banchio22a}, \citet{asker22}, \citet{banchio22b},
\citet{dolgopolov24} and \citet{hansen21}.\footnote{The more general
underlying economic motivation is whether the use of algorithms fosters
supra-competitive prices, a question that has also been examined by
\citet{brown21} without focusing specifically on reinforcement learning
algorithms.} Except for the later paper, which examines tacit collusion
induced by the use of UCB-learning (\citet{auer02}),\footnote{UCB stands for
uniform confidence bound. It builds an independent tract record for each
alternative (unlike $Q-$learning, which computes $Q(a)$ using $\max Q$ hence
the values $Q(a^{\prime})$ computed for actions different from $a)$.} the
others consider, like us, Q-learning algorithms (\citet{watkins92}).

We have already emphasized three important departures from these studies: (i)
we provide players with discretion over a set of biased policy rules; (ii) we
examine whether cooperation can be sustained even when
starting from initially unfavorable conditions; (iii) we allow for payoffs
being subject to common (as in \citet{calvano21}) or private shocks.

There are other differences. \citet{calvano20,calvano21} assume vanishing
experimentation and memory-based $Q-$values. Memory is important in the latter
work because when experimentation vanishes, memoryless Q-values are
ineffective in supporting collusion. Supporting collusion then requires some
conditioning on others' prices.\footnote{\citet{klein21} examines a similar
duopoly setup. There is no memory but moves are sequential, so effectively,
Q-values are assumed to be contingent on the rival's price.} In contrast,
\citet{banchio22a} consider experimentation that does not vanish and, under
some conditions, obtain cooperation even with memoryless Q-values. One
difference with our work is that BM characterizes the \textit{continuous
limit} of the $Q-$value updates when each player independently updates at
randomly distributed times (according to a Poisson distribution), which we do
not. Under this continuous limit, BM find that exit from bad initial
conditions is impossible, so if, for whatever reason, current payoff
conditions call for prolonged defection, players will be unable to eventually
recoordinate on cooperation.\footnote{The reason is that under the continuous
limit examined, the feedback from experimentation is instantaneous and this
precludes simultaneous experimentation. Keeping the continuous time
assumption, alternative assumptions about the duration of experimentation
necessary to obtain reliable feedback would yield different conclusions
regarding the possibility of recoordination (as sufficiently synchronous
experiments could then arise).}

Memoryless Q-learning algorithms have also being examined through the angle of
stochastic stability (\citet{foster90}), studying the limit behavior under
Q-learning for arbitrarily small experimentation probabilities
(\citet{waltman07,waltman08} and \citet{dolgopolov24}). In the absence of
experimentation, there are two stable outcomes, a cooperative one where
$Q_{i}(C)>Q_{i}(D)$ for both players, and a non-cooperative where
$Q_{i}(D)>Q_{i}(C)$ for both players. Exits from the \textquotedblleft
cooperative basin\textquotedblright\ arise out of consecutive one-sided
experimentation, while exits from the \textquotedblleft non-cooperative
basin\textquotedblright\ require consecutive \textit{joint} experimentation.
As \citet{dolgopolov24} shows, this asymmetry implies that under greedy
algorithms, only the non-cooperative outcome survives at the limit of
arbitrarily small experimentation.
He also observes that with Bolzmann experimentation and arbitrarily small
experimentation probabilities, one may generate arbitrarily larger chances of
experimentation in non-cooperative phases (compared to chances of
experimentation in the cooperative phase), and this asymmetry is conducive to
cooperation.\footnote{\citet{waltman07,waltman08} also use Bolzmann
exploration and exploit this same idea of asymmetric exploration probabilities
between cooperative and non-cooperative phase to generate collusion when
exploration probabilities are arbitrarily small.}

In contrast, in our work, experimentation is not arbitrarily small, and in the
noisy signals extension, noise actually precludes the possibility of staying
for ever in a given phase. Players transit from one phase to another, and we
obtain collusion and cooperation because biases modify transition
probabilities between phases, positive biases facilitating exit from
non-cooperative phases.

\textbf{Other related work.} Our work is also related to
\citeauthor{compte15}~(\citeyear{compte15} and \citeyear{compte18},
\href{http://www.parisschoolofeconomics.com/compte-olivier/Chapter14Cooperation.pdf}{\color{dark-blue}Ch.
14}) who examine repeated prisoner's dilemma with stochastic signals or
payoffs, where players are constrained to choosing a strategy within
a\ limited family of stochastic automata. The automata perform reasonably well
for a class of bandit problems where the risky arm has state-dependent
benefits and where the state has some persistence: once in a while, it pays to
check whether the risky arm turns out to be profitable again. The
\textit{frequency} with which one checks is then conceived as a strategic
variable. In the context of the prisoners' dilemma, cooperation is the risky
arm, and higher frequency of checking whether cooperation is profitable means
more leniency. Like here, lenient strategies facilitate recoordination on
cooperation, and CP find equilibrium leniency levels.\smallskip



Our work is also related to \textbf{evolutionary game theory}. This literature
investigates which strategies survive in the long-run, within a (possibly
large -- but nevertheless restricted) ecology of behavioral rules. In the
context of the prisoner's dilemma (\citet{axelrod84}), the behavioral rules
typically consist of deterministic or stochastic functions of recent action
profiles played (as in \citet{nowak90} and \citet{kraines00} for example).

One relevant and interesting exception is \citet{moriyama11} who consider
strategies based on \textit{modified Q-values}: while standard Q-values build
on the rewards $r$ actually received, \citet{moriyama11} define Q-values built
from subjectively modified rewards $u(r)$, where $u$ is a utility function
(not necessarily monotonic in $r$).\footnote{In the prisoner's dilemma with
perfectly observable outcomes, there are four possible rewards -- one for each
action profile $a=CC,CD,DC,DD$, so in effect, there are four possible degrees
of freedom for choosing the subjective rewards $u(r(a))$.} In each period,
these modified Q-values determine which action is best. \citet{moriyama11}
then simulate which utility function survive in the long run, based on fitness
(i.e., ex ante expected gains determined by actual rewards).\footnote{The
metholodogy described here is in the spirit of the literature on the evolution
of preferences: the agent use (modified) Q-values as proxies for continuation
values, and preferences $u(r)$ over rewards are endogenized to maximize
fitness (i.e., to maximize ex ante expected gains). See also \citet{singh10}.}
They find that evolution gives rise to (essentially) two utility levels, each
based on the \textit{other player's action only}. The connection with our work
is that technically, each utility function $u$ determines a strategy for the
entire game, so one can interpret their work as an endogenization of a
behavioral rule, within a parameterized family of behavioral rules. It remains
to be seen how their method would extend to stochastic payoffs and/or more
complex payoff structure with more than two actions, and whether the method
would help exit from defection traps in general.

Compared to the classic class of strategies examined in the literature cited
above, one benefit of working with our family of Q-based algorithms with
biased decision rules is that it can
be used across various payoff and monitoring environments, with the
finding that in all these environments, biased policies incorporate a simple
(one-dimensional) notion of \textit{leniency} that proves effective to support
cooperation or collusion. Leniency is of course at the heart of many
memory-one strategies like Tit-for-Tat, Pavlov or their stochastic variations
studied in the evolutionary literature, but examining evolutionary stability
within this space has proved difficult,
and these strategies are difficult to generalize to games with imperfect
monitoring.

Our work departs from most of the \textbf{bounded rationality literature},
which like us, aims to introduce limits on sophistication, but often does so
by introducing misspecifications (\citet{esponda16}), biased interpretation of
the environment or feedbacks (\citet{jehiel05}), or biased estimates of the
alternatives (\citet{osborne98}), with players maximizing a subjective
criterion which may not be congruent with their own true welfare. Here, one
may interpret the dated Q-value $Q_{i}^{t}(a)$ as a misspecified estimate of
the continuation value associated with playing $a$ at $t$, and the standard
(naive) Q-learning rule then corresponds to subjective maximization over
possible $a$'s at each date. But, when we examine Qb-equilibria and the
optimal bias for each player, we use the \textit{objective performance}
associated with bias profiles (measured by ex ante long-run payoffs).
The limit on sophistication comes from the (assumed) inability to evaluate all
strategies: only some strategies are considered, not all of them.

Finally, another related paper is \citet{karandikar98}, who consider players
who hold a slow-moving aspiration level (which geometrically aggregates own
past gains), and independently switch action with some probability when the
current gain is below the aspiration. There is no experimentation but
\textit{exogenous shocks} on the aspiration level lead to exogenous changes of
actions. \citet{karandikar98} show that when aspirations are (sufficiently)-
slow moving, convergence to CC is much easier than exiting from CC:
cooperation is the only stable outcome in the long-run. Intuitively, starting
from CC and aspiration levels above the value of mutual defection, DC creates
a disappointment for player 2, eventually leading to DD, which in turn
creates, since aspiration are only moving slow, joint disappointment and
repeated pressures back to CC. Inversely, starting from DD and low aspiration
levels, a shock on say player 1's aspiration level induces many CD's and DD's,
which raises the aspiration level of player 2 as well, hence eventually, as
explained above, repeated pressures back to CC whenever DD is played.

The mechanism by which cooperation is sustained in \citet{karandikar98} bears
some similarity with that induced by $Q$-learning. One can view $Q$-values as
\textit{action-specific} aspiration levels.
Cooperation is resilient under $Q$-learning because actions that are not
currently $Q$-optimal have only slow-changing aspiration levels: whenever
defection becomes optimal for say player 1, her $Q$-value associated with
cooperation only moves slowly;\footnote{It only changes when cooperation is
experimented. The role of these differentiated speeds of adjustment has been
emphasized by \citet{asker22} and \citet{banchio22a}. Competitive outcomes are
more likely when Q-learning is synchronous (as opposed to asynchronous). When
synchronous, players use observations to make inferences about the payoff they
would have obtained had they played differently, using these inferences to
update all $Q$-values at once.} the defection is quickly matched by player 2
and when this happens, player 1's $Q$-value of defection drops so cooperation
becomes attractive again for her.
One difference with \citeauthor{karandikar98} is that there is no easy way to
exit from DD under $Q$-learning, because CD just raises the Q-value of
defection for player 2, hence raises the barrier to cooperation for player 2,
while in \citeauthor{karandikar98}, CD raises the aspiration level of player
2, facilitating a subsequent switch to cooperation for player 2.

\section{Q-traps\label{sectionQtraps}}

We consider a single agent problem where payoffs are \textit{state-contingent}
and where the evolution of the state is not observable and partially governed
by the action played. In such environments, it is well-known that $Q$-learning
may be suboptimal (see \citet{singh94} or more recently \citet{barfuss22}).
Our example illustrates that, even when facing a relatively simple stochastic
Tit-for-Tat automaton, a Q-learner may fail to approach the maximum feasible
gain, whatever the exploration level $\varepsilon$\ and speed of adjustment
$\alpha$\ assumed.\footnote{The example contrasts with (but does not
contradict) \citet{sandholm96}, who shows that a more elaborate $Q$-learner
who build Q-values \textit{conditional on the last action profile played}
learns to play optimally against a Tit-for-tat player.} It also illustrates
that, even with small biases, Q-based learning can significantly improve welfare.

\textbf{The automaton.} The agent has two actions available, $a\in A=\{1,2\}$
and faces an automaton with two states and stationary transition probabilities
contingent on the current action of the agent. We let $\theta\in\{1,2\}$
denote the state and $\pi_{\theta\theta^{\prime}}^{a}=\Pr_{a}(\theta
\rightarrow\theta^{\prime})$ the probability to transit from $\theta$ to
$\theta^{\prime}$ when the agent plays $a$, and to fix ideas, we assume%
\[%
\begin{tabular}
[c]{c|cc}%
$\pi_{\theta\theta^{\prime}}^{a}$ & $1\rightarrow2$ & $2\rightarrow1$\\\hline
$a=1$ & $0.01$ & $0.05$\\
$a=2$ & $0.05$ & $0$%
\end{tabular}
\ \ \ \
\]
These transitions correspond to a stochastic TIT-for-TAT: interpreting action
1 as cooperation and action $2$ as defection, the automaton switches to the
(bad) state 2 with a small probability\ ($0.01)$ when facing an agent that
cooperates, and a larger probability $(0.05)$ when the agent defects. Once in
state 2, the automaton switches back to (good) state 1 with positive
probability $(0.05)$, but only when the agent cooperates\textit{.}

\smallskip

\textbf{Payoff structure.} We assume that payoffs are given by
\[%
\begin{tabular}
[c]{c|cc}%
$a\backslash\theta$ & $1$ & $2$\\\hline
$1$ & $2$ & $y$\\
$2$ & $x$ & $1$%
\end{tabular}
\ \ \
\]
where $y\in(-5,1)$ and $\frac{3x}{2}+y<5.$ The first condition ensures that in
state 2, action 2 is preferable in the short run. The first and second
conditions ensure that, \textit{even when the state is observable}, playing
$a=1$ at all dates is optimal. Furthermore, in simulations below, we focus on
the two cases $x=1$ and $x=2.5$.

Intuitively, the stochastic process alternates between \textquotedblleft
good\textquotedblright\ phases (where $\theta=1$) and \textquotedblleft
bad\textquotedblright\ phases (where state $\theta=2$). Playing defection in
state 2 is suboptimal because it keeps the process in a bad phase. Playing
defection in state 1 (and cooperation in state 2 to come back to state 1) is
suboptimal because even when $x>2$, it takes time to come back to the good
state, and overall, under the above conditions, cooperating always is
preferable. This strategy induces a long-run distribution over states which
puts weight $q^{0}=5/6$ on state 1, hence an expected gain
\[
u^{0}\equiv2q^{0}+(1-q^{0})y
\]
It will also be convenient to refer to $u^{\varepsilon}$ as the long-run
payoff obtained by the agent who plays $a=1$ with probability $1-\varepsilon$
in every period.\footnote{The induced distribution over states then puts a
weight $q^{\varepsilon}=5(1-\varepsilon)/(6-\varepsilon)<q^{0}$ on state $1$,
and we have $u^{\varepsilon}=(2(1-\varepsilon)+\varepsilon x)q^{\varepsilon
}+(y(1-\varepsilon)+\varepsilon)(1-q^{\varepsilon})<u^{0}$.}

\textbf{Q-learning.} We now consider a \textquotedblleft naive" $Q-$learner
who revises $Q^{t}(a)$ for $a\in\{1,2\}$ according to (\ref{Q}),
experiments with probability $\varepsilon$ in any period, and otherwise
follows the naive policy, i.e., chooses action 1 whenever
\[
\Delta^{t}\equiv Q^{t}(1)-Q^{t}(2)\geq0
\]
At any point in time, the dynamics is fully determined by $\Delta^{t}$ and
$\overline{Q}^{t}=\max_{a}Q^{t}(a)$. Because of experimentation, the maximum
feasible gain $u^{0}$ cannot be attained. But in principle, if $\Delta^{t}$
remained everywhere positive, the agent could secure $u^{\varepsilon}$.
However, Q-learning fails to approach $u^{\varepsilon}$: action $a=2$ is
played much too often, tilting the distribution over states towards $\theta
=2$. There are two reasons for that.

First, the agent may be trapped playing $(a,\theta)=(2,2)$ for long durations.
When $\overline{Q}^{t}\leq1$ and $\Delta^{t}<0$), defections keeps the state
in state $2$, and occasional experimentation confirms that cooperation is
worse (because $y<1$). Escaping the trap is possible with enough
experimentation and/or high enough adjustment speed $\alpha$, but whatever
facilitates exits from the trap (i.e., high $\alpha$ and $\varepsilon$) also
facilitates adjustments away from the desirable long-run outcome
$(a,\theta)=(1,1)$ --back to the trap.

Second, even within high-Q phases (where cooperation is more prevalent),
defection must be played a substantial fraction of the time: inevitably, the
state eventually transits to a bad phase ($\theta=2$).
During this bad phase, gains are below Q-values, so the Q-value \textit{of the
action played} must decrease, which results in frequent alternations between
the actions played, hence a significant weight on action 2.

\begin{figure}[h]
\centering
\begin{minipage}{0.45\textwidth}
\centering
\includegraphics[scale=0.6]{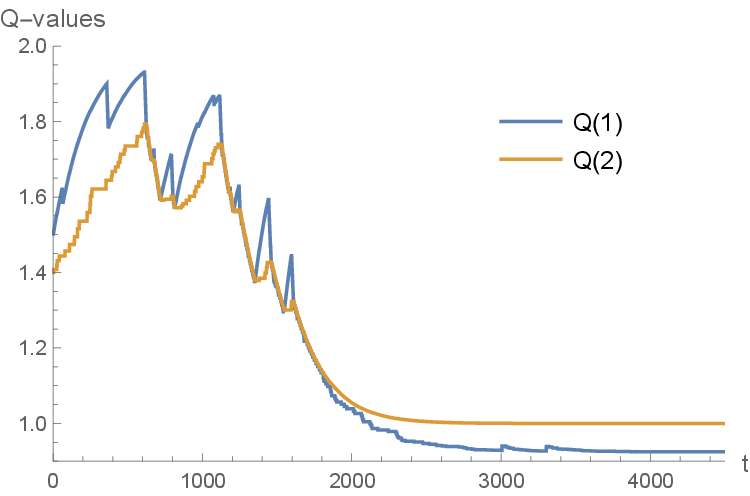}
\subcaption{Evolution of Q-values}
	\label{fig1}
\end{minipage}\hfill\begin{minipage}{0.45\textwidth}
\centering
	 \includegraphics[scale=0.6]{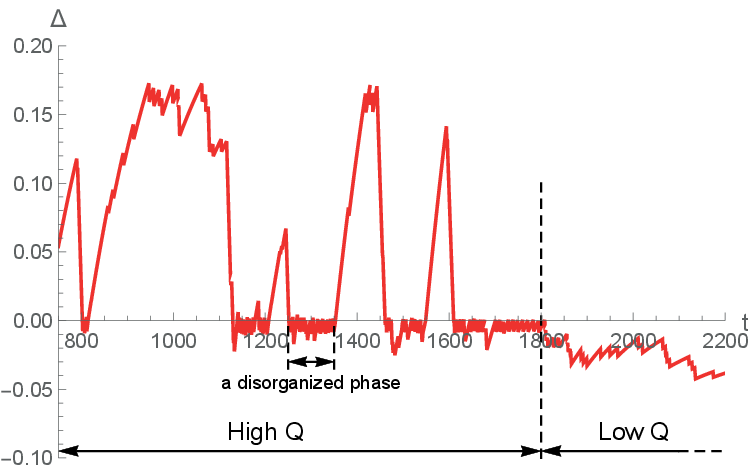}
\subcaption{Evolution of $\Delta$}
\label{fig2}
\end{minipage}
\caption{Q-learning Dynamics with $\alpha=0.1$ and $\varepsilon=0.1$}%
\end{figure}

\textbf{An illustration.} Figure \ref{fig1} describes the dynamics of Q-values
when $x=1$ and $y=-0.5$ under low speed and low experimentation $(\alpha=0.1$
and $\varepsilon=0.1$), starting from favorable initial conditions
($Q^{0}(1)=1.5$, $Q^{0}(2)=1.4$ and $\theta=1$). For the path considered, the agent keeps staying in the favorable
state during about 600 periods, until the state changes durably enough to
$\theta=2$ so that $Q^{t}(1)$ drops below $Q^{t}(2)$, making action $a=2$ more
attractive than $a=1$.

Then follows what we shall call a \textit{disorganized phase} where $\theta=2$
and $\Delta^{t}$ remains small. In that phase, the agent quickly alternates
between his two actions. The alternation is not triggered by experimentation,
but the fact that under $\theta=2$, current payoffs are below both $Q$-values.
The Q-value of the action played thus decreases and the action not played
becomes more attractive. So long as $\theta=2$, $\Delta^{t}$ remains small and
its sign is repeatedly changing, as illustrated in Figure~\ref{fig2}. This
phase may stop after a transition to $\theta=1$: if $a=1$ is played
repeatedly, the $Q$-value of action $1$ surges, fostering a prolonged
favorable phase where $\overline{Q}$ rises again.

\begin{figure}[h]
\centering
\includegraphics[scale=0.65]{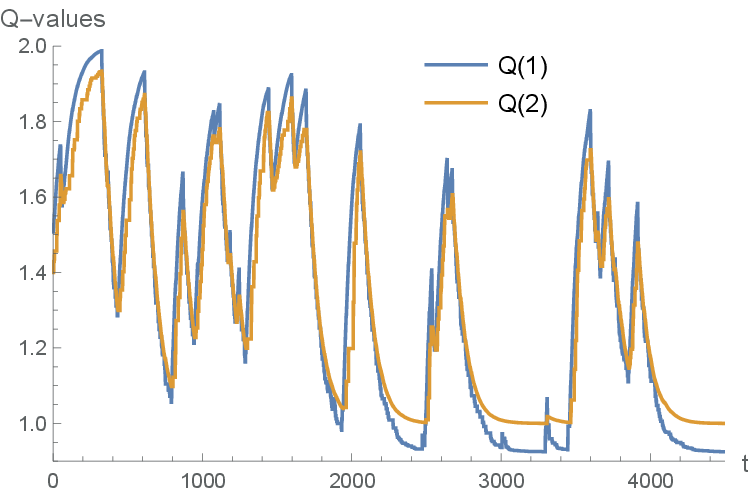}\caption{Evolution of $Q$-values with
$\epsilon=0.3$ and $\alpha=0.3$}%
\label{Fig3}%
\end{figure}

However, it may also happen that the state $\theta=2$ is persistent enough to
drive $Q^{t}(1)$ below $1$, and a low-Q phase starts (see Figure~\ref{fig2}).
When this happens, occasional experimentation does not help: it just confirms
that $Q^{t}(1)$ is a worse action, and the agent remains trapped playing $a=2$
for a long duration. Escaping the Q-trap is possible with enough
experimentation and large enough speed of adjustment $\alpha$ (see
Figure~\ref{Fig3}). However, if high $\varepsilon$ and $\alpha$ facilitate
exits from traps, they also induce fast adaptation whenever the state changes
back to $\theta=2$, so exits are generally not long-lasting.

To confirm this intuition, we performed simulations over a very
long horizon $\overline{T}=800000$ periods, starting from unfavorable
conditions (i.e., $Q(1)=0.9<Q(2)=1$), for all pairs $(\alpha,\varepsilon
)=(k\alpha_{0},k^{\prime}\varepsilon_{0})$ for $\alpha_{0}=0.05$,
$\varepsilon_{0}=0.03$, with $k,k^{\prime}\in\{1,..,9\}$. For $(x,y)=(1,0)$,
we find that across all pairs $(\alpha,\varepsilon)$, the average frequency of
$(a,\theta)=(1,1)$ is bounded above by $0.3$ and the average frequency of
$a=2$, conditional on $\theta=2$, is bounded below by 0.78. When
$(x,y)=(2.5,0.5)$, these average frequencies are respectively $0.08$ and
$0.77$. So in both cases, we are quite far from efficiency (which requires
choosing $a=1$ conditional on $\theta=2$.)

\textbf{Comparative statics and the effect of biased Q-learning}. Next we fix
$(\alpha,\varepsilon)=(0.1,0.1)$ and report welfare levels obtained over the
very long horizon $\overline{T}$ (long enough that given $(\alpha
,\varepsilon)$, initial conditions have negligible effects), as we vary the
payoff parameters $x$ and $y$. We also indicate the welfare levels obtained
conditional on $\overline{Q}>1.2$. This allows us to disentangle the two
sources of losses mentioned above, the one stemming from the use of action 2
in high-Q phase and the one stemming from being trapped in a low-Q phase. We
also report the welfare levels obtained under a Q-based automaton with a small bias.

\begin{figure}[h]
\centering
\begin{minipage}{0.45\textwidth}
\centering
\includegraphics[scale=0.5]{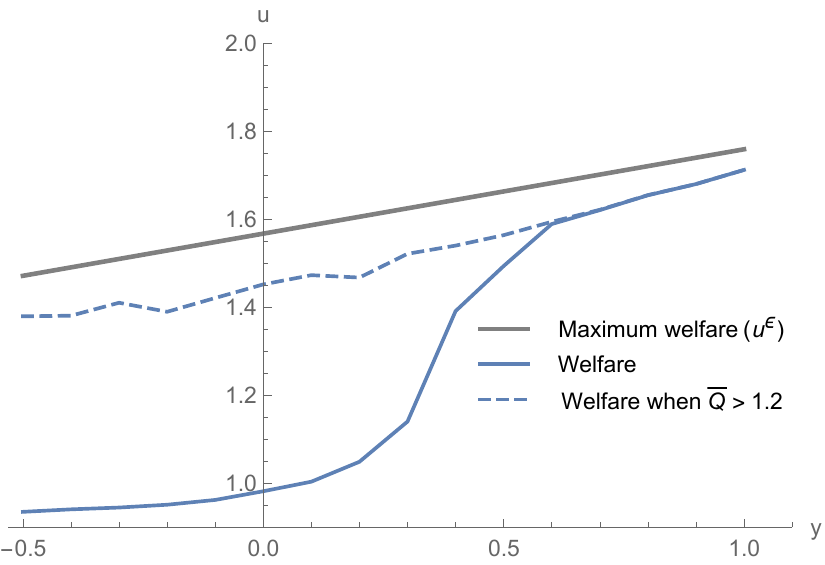}
\subcaption{$b=0$}
	\label{figdec}
\end{minipage}\hfill\begin{minipage}{0.45\textwidth}
\centering
	 \includegraphics[scale=0.5]{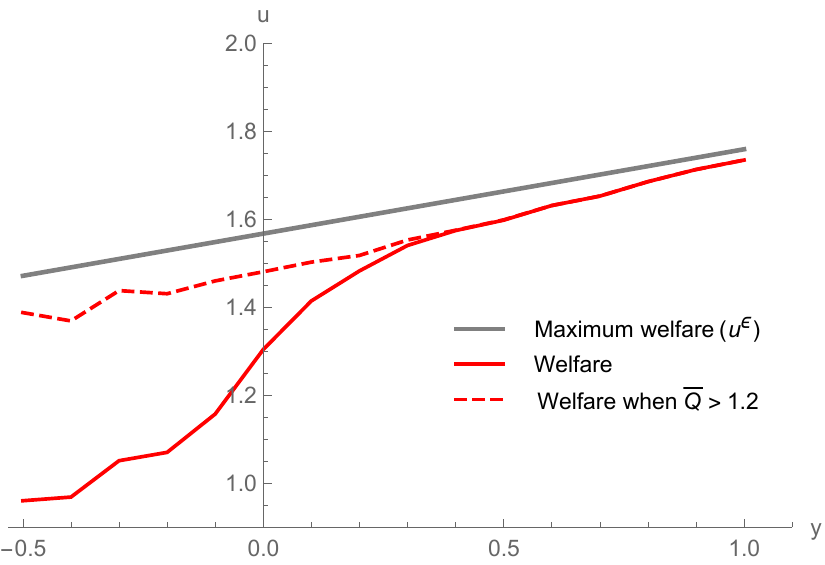}
\subcaption{$b=0.02$}
\label{figdecbiased}
\end{minipage}
\caption{Welfare levels, $x=1$}%
\end{figure}

First set $x=1$. We observe three domains as we vary $y$ (see
Table~\ref{figdec}). When $y$ is too low ($y\leq0.2$), exiting from a Q-trap
is hard and long-run payoffs remain close to 1. When $y$ is high enough
($y\geq0.6$ for $x=1$), high-Q phases prevail. For intermediate values of $y$,
the agent alternates between high-Q and low-Q phases. Conditional on
$\overline{Q}>1.2$, the loss is approximately independent of $y$. So losses
vary with $y$ mostly because of the effect on the persistence/prevalence of Q-traps.

Figure~\ref{figdecbiased} reports welfare levels when the agent's decision
rule is biased, with $b=0.02$. The bias does not affect much the dynamics
within high-Q phases: disorganized phases now occur when $\theta=2$ and
$\Delta^{t}\equiv Q^{t}(1)-Q^{t}(2)+b$ is close to 0. The bias essentially
diminishes the duration of low-Q phases and their prevalence in the long-run.

\begin{figure}[h]
\centering
\begin{minipage}{0.45\textwidth}
\centering
\includegraphics[scale=0.5]{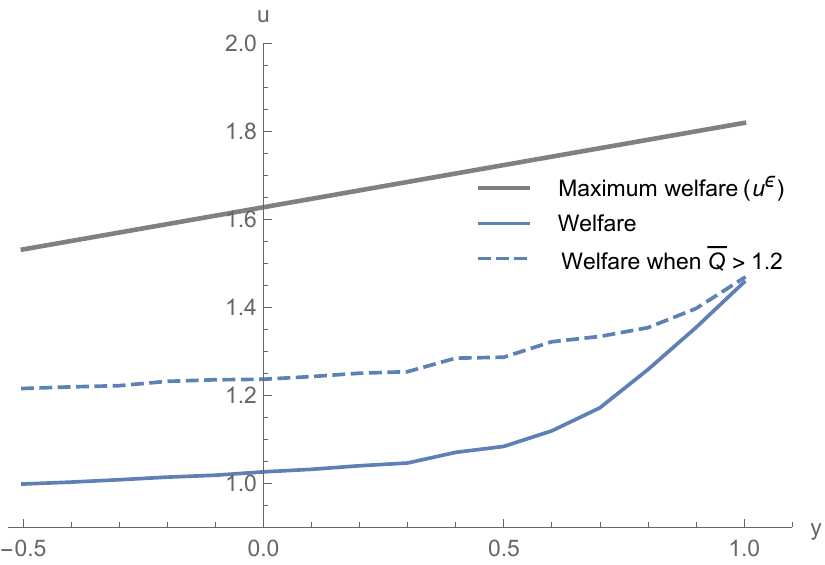}
\subcaption{$b=0$}
	\label{figdecx25}
\end{minipage}\hfill\begin{minipage}{0.45\textwidth}
\centering
	 \includegraphics[scale=0.5]{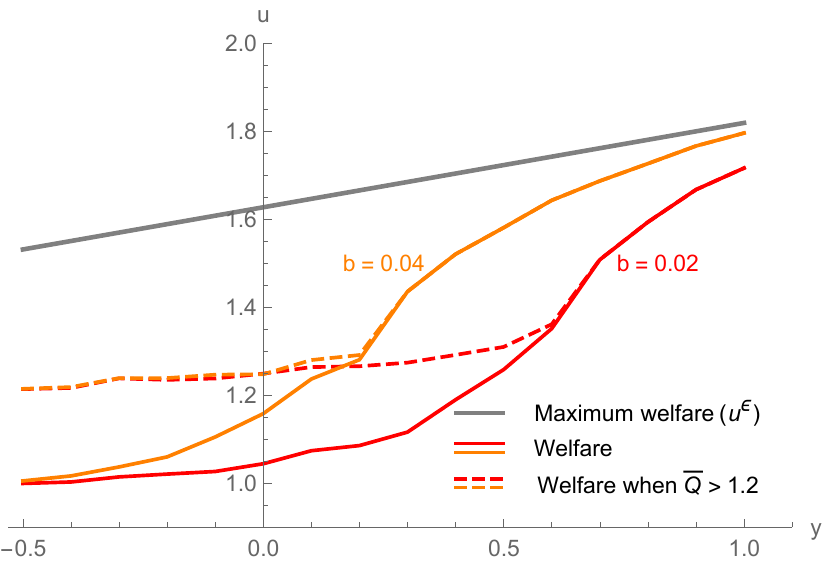}
\subcaption{$b=0.02$ and $b=0.04$}
\label{figdecbiasedx25}
\end{minipage}
\caption{Welfare levels, $x=2.5$}%
\end{figure}

When $x=2.5$, we obtain similar dynamics
with alternation between high-Q and low-Q phases. One difference is that
experimentation with $a=2$ always favors that action (because $x>2$ and
$1>y$). This implies that exits from traps are more difficult. But it also
implies that even within high-Q phases, $\Delta^{t}$ is often negative (so
$a=2$), and this tilts the distribution over states further away from the
efficient level, as Figure~\ref{figdecx25} confirms. We nevertheless obtain a
strong effect of biases on welfare even for small biases (see
Figure~\ref{figdecbiasedx25}) 

Finally, let us comment on the magnitude of the bias sufficient to facilitate
exits from traps. On average, within a Q-trap, $\Delta$ is on average equal to
$\overline{\Delta}=(1-\delta)(1-y)$.\footnote{The reason is that when
experimentation occurs, $Q(1)$ is updated using $\overline{Q}$} Overcoming
this gap can be hard when $\alpha$ is small because it may require several
periods of consecutive experimentation. A bias $b$ equal to say $\overline
{\Delta}/2$ has a significant effect the probability of exit because it
reduces by a factor 2 the number of consecutive experimentations necessary.

\section{The Prisoner's Dilemma\label{section3}}

We now examine the interaction of two $Q$-learners, each trying to determine
whether to cooperate or defect based on the $Q$-values of cooperation and
defection. Payoffs are given by:%
\[%
\begin{tabular}
[c]{c|cc}%
$a_{1}\backslash a_{2}$ & $C$ & $D$\\\hline
$C$ & $2$ & $y$\\
$D$ & $x$ & $1$%
\end{tabular}
\
\]
In simulations below we set $x=2.5$ and $y=-0.5$. We perform comparative
statics with respect to $y\in(-1,1)$ in Section~\ref{seccomp}. Note that when
$x+y=2$, only CC yields a Pareto superior payoff, so joint expected gains
exactly reflect the extent of joint cooperation. Define $\overline{Q}_{i}%
^{t}\equiv\max_{a_{i}\in\{C,D\}}Q_{i}(a_{i})$ and
\[
\Delta_{i}^{t}=Q_{i}^{t}(C)-Q_{i}^{t}(D)\text{ and }\rho_{i}^{t}=\Delta
_{i}^{t}-\Delta_{i}^{t-1}.
\]

\subsection{Naive Q-learning}

The dynamics of naive Q-learning is well understood, but we find instructive
to clarify here the typical phases that players undergo and to describe the
transitions between phases. This is the purpose of this Section.

First note that while players do not face a simple two-state automaton, their
behavior bears some resemblance with it: confronted with a player that mostly
defects, occasional experimentation will confirm that $D$ is the best option,
and be conducive to further defection. As a result players may be trapped in
$DD$ for some time.

Exiting DD is possible but difficult. It requires \textit{simultaneous}
experimentation (hence sufficiently frequent experimentation on each side), so
as to induce some $CC$'s which, if adaptation $\alpha$ is not too small, will
induce a simultaneous rise of $\Delta^{t}$ for both that may propel the
interaction into a cooperative phase. Unfortunately, while high
experimentation levels and/or high adaptation allow exits, they also hurt the
sustainability of cooperation, as in our previous automaton example. Table
\ref{T4} reports expected welfare levels over a 10.000-period horizon,
starting from unfavorable conditions, for different values of $\alpha$ and
$\varepsilon$, assuming $x=2.5$ and $y=-0.5.$\footnote{For $y=-0.5$, the
welfare above 1 coincides with the fraction of the time players cooperate.
Payoffs are obtained by running 90 simulations of 10000 periods, starting from
$Q_{i}(C)=0.95$ and $Q_{i}(D)=1$ for both players, and computing the mean
payoff obtained.} \begin{table}[h]
\caption{Naive Q-learning}
\begin{subtable}{.5\linewidth}
\centering
\subcaption{Welfare levels}
\label{T4}
\scalebox{0.7}{
\begin{tabular}[c]{c|cccccccc}
\text{$a\backslash \epsilon $} & 0.025 & 0.05 & 0.075 & 0.1 & 0.125 & 0.15 & 0.175 & 0.2 \\\hline
0.1 & 1. & 1. & 1. & 1. & 1. & 1.01 & 1.01 & 1.01 \\
0.3 & 1. & 1.03 & 1.03 & 1.06 & 1.06 & 1.04 & 1.04 & 1.04 \\
0.5 & 1.02 & 1.11 & 1.18 & 1.16 & 1.16 & 1.1 & 1.08 & 1.07 \\
\end{tabular}}
\end{subtable} \begin{subtable}{.5\linewidth}
\centering
\subcaption{Chances of exit}
\label{T4b}
\scalebox{0.7}{
\begin{tabular}[c]{c|cccccccc}
\text{$a\backslash \epsilon $} & 0.025 & 0.05 & 0.075 & 0.1 & 0.125 & 0.15 & 0.175 & 0.2 \\\hline
0.1 & 0. & 0. & 0. & 0. & 0. & 0. & 0. & 0. \\
0.3 & 0. & 0.06 & 0.1 & 0.2 & 0.29 & 0.47 & 0.63 & 0.86 \\
0.5 & 0.04 & 0.27 & 0.67 & 0.92 & 0.98 & 1. & 1. & 1. \\
\end{tabular}}
\end{subtable}
\end{table}
We also report the chances of exit from defection over this 10.000-period
horizon (Table~\ref{T4b}).

Chances of exit are significant when $\alpha=0.5$ and $\epsilon\leq0.05$, but
welfare remains low because cooperation does not persist.

\begin{figure}[h]
\centering
\includegraphics[scale=0.6]{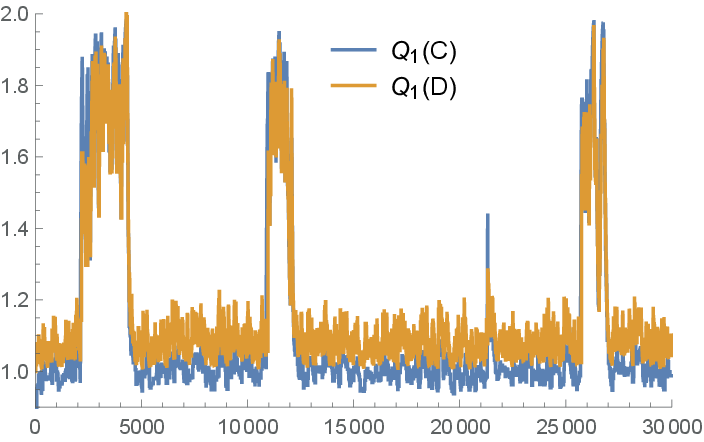}\caption{Evolution of $Q$-values for
$\alpha=0.5$ and $\varepsilon=0.1$}%
\label{fig6}%
\end{figure}

\textbf{The dynamics of Q-values}. To illustrate the dynamics of Q-values,
consider a path of Q-values obtained when setting $\alpha=0.5$ and
$\varepsilon=0.1$ (see Figure~\ref{fig6}). For these $\varepsilon$ and
$\alpha$, we see an alternation between high-Q phases where cooperation occurs
and low-Q phases where players mostly defects. Exits from low-Q phases are
possible (over the time horizon shown), but high-Q phases do not last very
long, compared to low-Q phases.\footnote{We plot the Q-values of player 1.
$Q_{1}(C)$ is the blue curve, $Q_{2}(D)$ is the orange curve.}

The occasional peaks correspond to phases where cooperation occurs, but on a
closer look, these peaks involve relatively rapid alternations of
\textbf{jointly-cooperative phases} (where $\Delta_{i}^{t}>0$ for both) and
\textbf{disorganized phases} (composed mostly of $CD^{^{\prime}}$s and
$DD$'s). In \textbf{disorganized phases}, $Q$-values remain high and the
payoffs obtained are below $Q$-values on average. As in
Section~\ref{sectionQtraps}, this puts a downward pressure on $Q$-values, and
it also prompts players to switch action frequently (with both\ $\Delta
_{i}^{t}$ remaining close to $0$). This frequent switching of action is what
gives cooperation \textit{resilience}, because eventually both will likely
cooperate simultaneously and enjoy payoffs above Q-values, confirming that
cooperation is a good option and fueling a new rise in Q-values.

\begin{figure}[h]
\centering
\begin{minipage}{0.45\textwidth}
\centering
\includegraphics[scale=0.45]{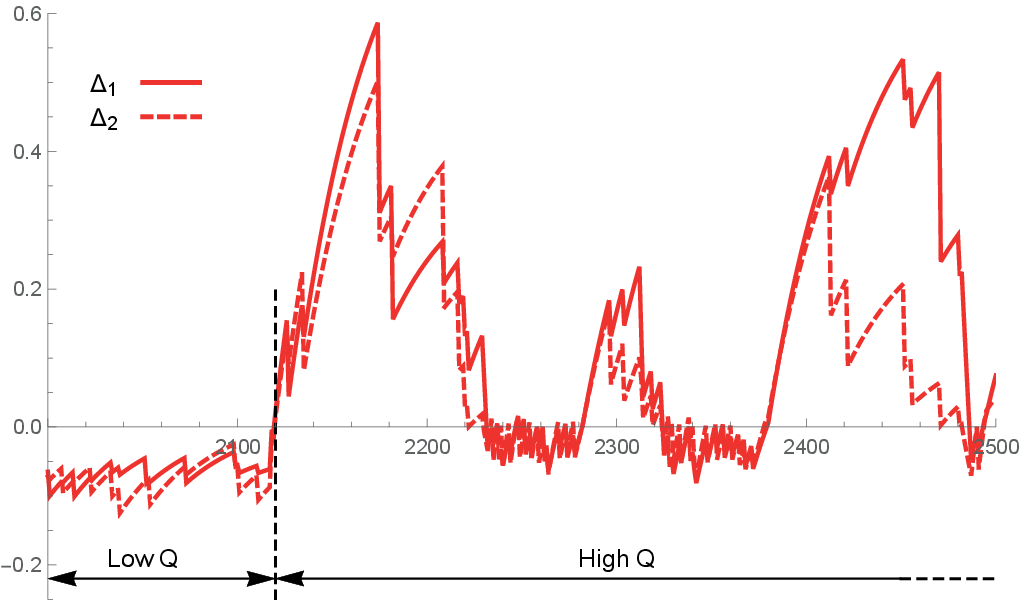}\subcaption{Exit from Q-trap}%
\label{fig7}
\end{minipage}\hfill\begin{minipage}{0.45\textwidth}
\centering
	\includegraphics[scale=0.45]{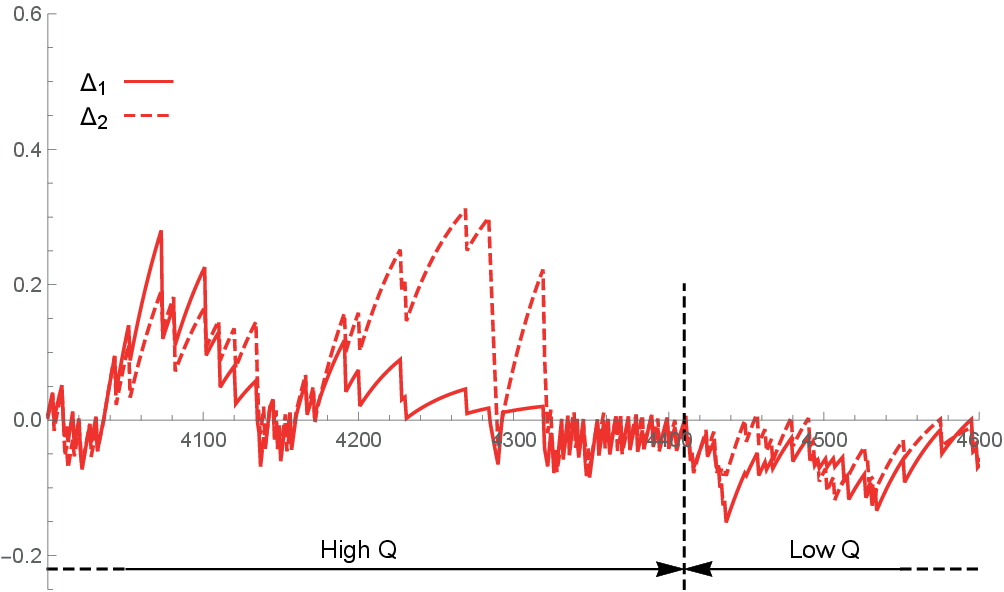}\subcaption{Back to defection}%
\label{fig8}
\end{minipage}
\caption{Transitions between phases}%
\end{figure}

\textbf{Transitions between phases.} Figure \ref{fig7} plots the dynamics of
$\Delta$'s between $t=2000$ and $t=2500$, starting from a defection trap (with
low $Q$'s). At $t=2120,$ players simultaneously jump to a prolonged
cooperative phase which increases durably $Q$-values. Within this high-Q
phase, alternations between jointly-cooperative phases (where $\Delta_{i}$ is
high for both) and disorganized phases (where $\Delta_{i}$ are both close to
$0$) occur.
Figure~\ref{fig8} shows the end of the high-Q phase, which terminates with a
prolonged disorganized phase. The duration of the disorganized phase is long
enough to induce a drop in $Q-$values that makes prospects of cooperation slim
($Q-$values drop below 1), and players are trapped again in a (long) defection
phase.

\textbf{The role of disorganized phases.} Consider a (relatively-)high
$Q-$phase, where $\overline{Q}_{i}^{t}\in(1,2)$. When either $CC$ or $DD$ is
played, both $\Delta_{i}$'s must rise (because $1<\overline{Q}_{i}^{t}$ so
playing $DD$ decreases the value of defection, and because $\overline{Q}%
_{i}^{t}<2$ so $CC$ increases the value of cooperation). When $CD$ or $DC$ is
played, both $\Delta_{i}$ must decrease. Thus, for any action profile
played,\textit{\ the signs of }$\rho_{1}$\textit{\ and }$\rho_{2}$ \textit{are
identical.} Magnitudes of $\rho_{1}$ and $\rho_{2}$ however differ, either
because $Q$-values do not coincide or whenever $CD$ is played.
%
This implies that in general, the $\Delta_{i}^{\prime}s$ do not cross the
$0-$frontier at the same time: so we must have $\Delta_{i}^{t}>0>\Delta
_{j}^{t}$ and a disorganized phase must start.

\begin{figure}[h]
\centering
\includegraphics[scale=0.6]{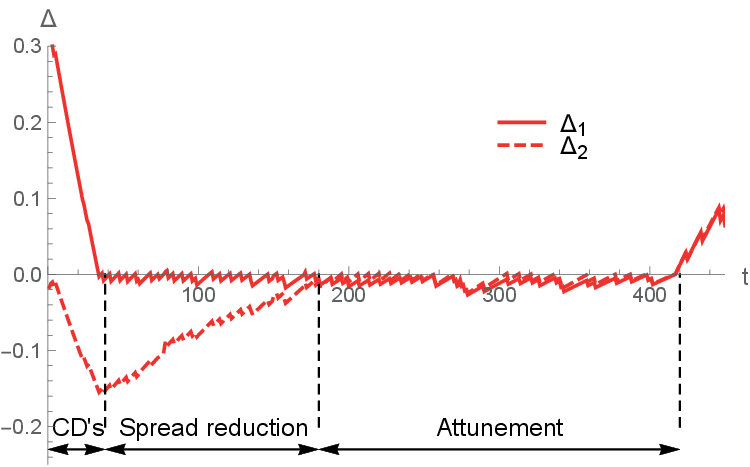}\caption{The anatomy of a disorganized phase}%
\label{f8}%
\end{figure}

Figure \ref{f8} illustrates one such instance, where we force initial
conditions to be sufficiently asymmetric.\footnote{We choose $\Delta_{1}=0.3$
and $\Delta_{2}=-0.01$ (with $\max Q_{1}=1.4$ and $\max Q_{2}=1.32$). We force
the initial asymmetry to be large to highlight the presence of the CD's and
spread reduction phases on the figure.} Disorganized phases consists of three sub-phases. First, CD's put
downward pressures on the $\Delta_{i}$'s, until $\Delta_{2}<\Delta_{1}<0$. At
this point, $\Delta_{1}$ is close to $0$ and a \textbf{spread-reduction phase}
starts where, unless experimentations occur, player 2 plays $D$ and player 1
alternates between $Cs^{\prime}$ and $Ds^{\prime}$,\footnote{Along this phase,
player 1 keeps alternating between C$^{\prime}s$ and $D$'s because $\Delta
_{1}$ remains close to 0 as $CD$'s put pressures downward while $DD$'s put
pressures upward.} until the point where both $\Delta$'s are close to $0$. In
the last phase, which (by chance) may be very short (but can sometimes be
long), players alternate between $CD$'s $DC$'s and $DD$'s until both
simultaneously switch to $\Delta_{i}>0$, with cooperation then driving up both
$\Delta$'s and $Q$'s. We call this phase an \textbf{attunement phase}.

Sometimes attunement can be very long and the disorganized phase then
terminates in a defection phase. Intuitively, only $DD$'s put pressures upward
on $\Delta$'s, but DD's necessarily reduce $Q$'s values. Furthermore, as $Q$
values become smaller, the upward pressure becomes weaker,\footnote{$\Delta
_{i}$ increases when $DD$ is played insofar as $\max Q_{i}>1$.} and
eventually, the $\Delta_{i}$'s remain persistently below 0.

In summary, the role of a disorganized phase is to reduce the spread
$|\Delta_{i}-\Delta_{j}|$ and keep both $\Delta$'s close to $0$, which allows
for the possibility of a simultaneous recoordination on cooperation.

\subsection{Q-based learning and equilibrium biases\label{sectionQbased}}

We now investigate the consequence of agents adopting biased policy rules,
whereby a player chooses to cooperate whenever%
\[
Q_{i}(C)+b_{i}>Q_{i}(D)
\]
A choice $b_{i}>0$ seems conducive to more cooperation for player $i$, but it
is not clear that this is a good policy: if $Q$-values are good proxies for
continuation values, a biased policy can only hurt; but worse, being more
cooperative might actually have the drawback of making defection more
attractive to one's opponent, hence to generate more joint defections
eventually. When Q-traps are an issue however, we will find that being more
lenient may help and foster joint cooperation beneficial to both.

Technically we consider a finite grid of possible biases $b_{i}\in B_{i}$, and
assume $b_{i}=\kappa_{i}\varpi$ for $\kappa_{i}\in\{\underline{\kappa
},...,\overline{\kappa}\}$, with $\underline{\kappa}\leq0<\overline{\kappa}$,
for some fixed increment $\varpi$, which we set at $\varpi=0.02$.\footnote{The
relevant magnitude is $\varpi/Q$. Since $Q$-values are normalized, and
typically equal to $1$ in a long defection phase, each increment corresponds
to a 2\% bias.} We choose the size of the increment to limit the number of
simulations. This forces a particular discretization of the space of biases,
but the effect on equilibrium outcomes turns out to be limited. A
discretization is also justified by the fact that we derive payoffs associated
with bias profiles using simulations. These simulations only give an
approximation of expected payoffs, so performance comparison between two very
nearby biases most likely reflect noise rather than true performance.

For a given $y$ and bias profile $(\kappa_{1},\kappa_{2})$, we simulate $n$
paths of play, each of length $T$ periods, starting from initially unfavorable
conditions (where $Q_{i}(C)=0.95$ and $Q_{i}(D)=1$ for both players). We
compute the average gains $v_{i}^{y}(\kappa_{i},\kappa_{j})$ obtained over
these paths.
Since the draws may by chance favor one player, we further anonymized payoffs
and compute $\overline{v}^{y}(\kappa,\kappa^{\prime})=(v_{i}^{y}(\kappa
,\kappa^{\prime})+v_{j}^{y}(\kappa,\kappa^{\prime}))/2$. Table~\ref{T6}
reports one such gain matrix $\overline{v}^{y}$, as well as, for each pair
$\kappa=(\kappa_{1},\kappa_{2})$, the fraction of the time joint cooperation
occurs.\footnote{Simulations in Table \ref{T6} are done with $n=90$,
$T=10000$, $\alpha=0.5,\delta=0.95$ and $\varepsilon=0.1$, chosen so that
players experience many alternations between all phases (traps, cooperative
and disorganized) over the short horizon considered here. Other simulations
are done over longer horizons (100000 periods).}

\begin{table}[h]
\caption{Biased Q-learning with $y=-0.5$}%
\label{T6}%
\begin{subtable}{.5\linewidth}
\centering
\caption{Gain matrix $\overline{v}^y(\kappa_{1},\kappa_{2})$}
\label{T6a}
\scalebox{0.85}{
\begin{tabular}[c]{c|ccccc}
$\kappa _1\backslash\kappa _2$ & 0 & 1 & 2 & 3 & 4 \\\hline
0 & 1.16 & 1.41 & 1.56 & 1.66 & 1.89 \\
1 & 1.38 & 1.61 & 1.71 & 1.77 & 1.84 \\
2 & 1.45 & 1.63 &\textbf{1.74} & 1.8 & 1.87 \\
3 & 1.4 & 1.57 & 1.69 &\textbf{1.81} & 1.88 \\
4 & 0.5 & 1.19 & 1.49 & 1.69 & 1.87 \\
\end{tabular}
}
\end{subtable} \begin{subtable}{.5\linewidth}
\centering
\caption{Frequencies of $CC$}
\scalebox{0.85}{
\begin{tabular}[c]{c|ccccc}
$\kappa _1\backslash\kappa _2$ & 0 & 1 & 2 & 3 & 4 \\\hline
0 & 0.16 & 0.39 & 0.5 & 0.53 & 0.19 \\
1 & 0.39 & 0.61 & 0.67 & 0.67 & 0.52 \\
2 & 0.5 & 0.67 & 0.74 & 0.74 & 0.68 \\
3 & 0.53 & 0.67 & 0.74 & 0.81 & 0.78 \\
4 & 0.19 & 0.52 & 0.68 & 0.78 & 0.87 \\
\end{tabular}
}
\end{subtable}
\end{table}

For example, $\overline{v}(2,0)=1.45$ provides the payoff of \textit{player 1}
when she biases her policy rule by $2\varpi=4\%$ while player 2 does not,
i.e., $\kappa=(2,0)$. The payoff of player 2 is obtained by looking at the
entry $\overline{v}(0,2)=1.56$. The frequency of cooperation for the profile
$\kappa=(2,0)$ is $50\%$. The bias $2\varpi$ by player 1 generates a payoff
asymmetry because the lenient policy of player~1 modifies the distribution of
CD's and DC's at her expense. Nevertheless, the effect on the occurrence of
cooperation is strong, as it increases from 16\% to 50\%, and this provides
each player with incentives to bias their policy away from the naive policies
$\kappa=(0,0)$.

From Table \ref{T6a}, it is immediate to check that there are two pure
equilibria $b^{\ast}=2\varpi$ and $b^{\ast}=3\varpi$ that both yield a payoff
significantly higher that what naive Q-learning delivers. Table \ref{T6a} also
shows that $\overline{v}^{y}(\kappa_{1},\kappa_{2})$ is rising in $\kappa_{2}%
$, so player 1 benefits from a stronger bias by his opponent benefits.
Interestingly however, the reason why player 1 benefits depends on the
magnitude of $\kappa_{2}$. A small rise away from $0$ benefits both player 1
and player 2 because this strongly increases the chance of joint cooperation.
For $\kappa_{2}>3$ however, the first line of Table \ref{T6}b shows that the
biased policy rule of player 2 \textit{decreases} the chance of joint
cooperation: it makes defections more attractive to player 1 and at
$\kappa_{2}=4$, the chance of joint cooperation drops down to 19\% -- player
\ 1 mostly benefits from player 2's generous policy by frequently defecting
(and enjoying 2.5): as Table~\ref{T8} shows, when the bias of a player is too
large, the distribution over action profiles played favors the agent with a
smaller bias.
%

\begin{figure}[h]
\centering
\begin{minipage}{0.45\textwidth}
\centering
	\includegraphics[scale=0.55]{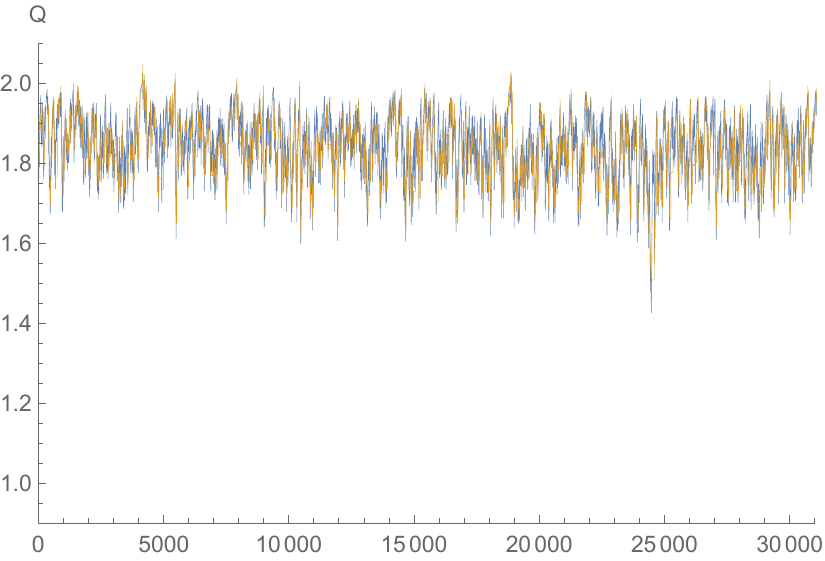}\subcaption{Long horizon}%
\label{fig10a}
\end{minipage}\hfill\begin{minipage}{0.45\textwidth}
\centering
\includegraphics[scale=0.55]{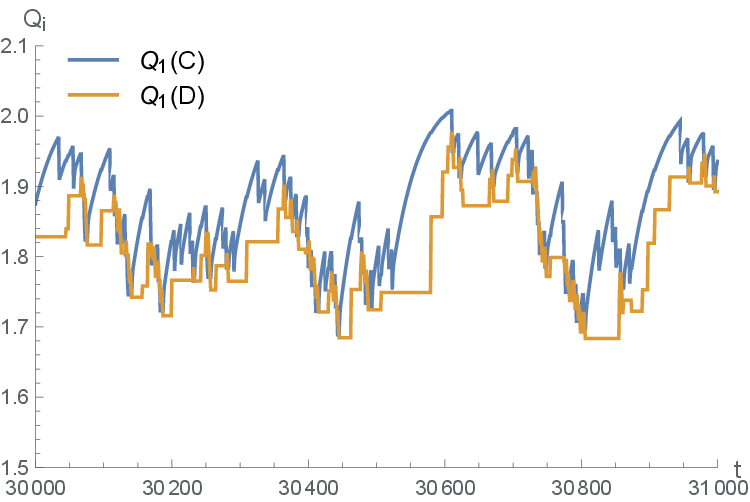}\subcaption{Short horizon}%
\label{fig10b}
\end{minipage}
\caption{Evolution of $Q$-values under biased $Q$-learning}%
\label{fig10}%
\end{figure}

\begin{figure}[h]
\centering
\includegraphics[scale=0.55]{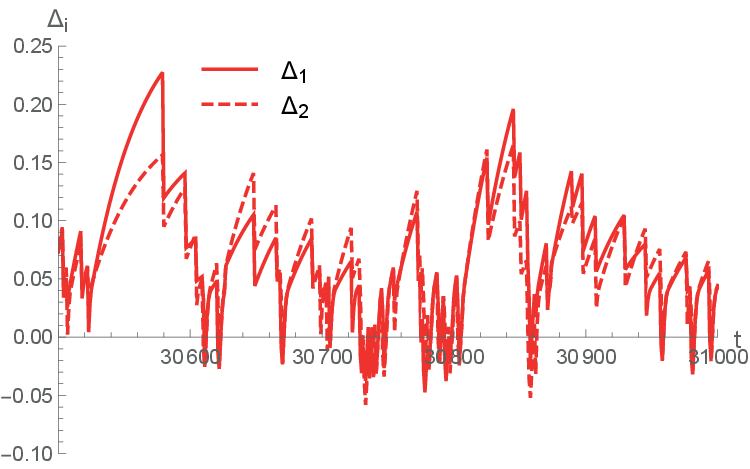}\caption{Evolution of $\Delta$ under
biased $Q$-learning}%
\label{F9}%
\end{figure}

\begin{table}[h]
\caption{$Pr(a|\kappa)$}%
\label{T8}%
\centering
\scalebox{0.85}{
\begin{tabular}
[c]{c|cccc}{\small $\kappa$}${\small \backslash\%}$ & ${\small CC}$ & ${\small CD}$ &
${\small DC}$ & ${\small DD}$\\\hline
${\small (0,0)}$ & ${\small 16}$ & ${\small 6}$ & ${\small 6}$ & ${\small 73}$\\
${\small (2,0)}$ & ${\small 50}$ & ${\small 9}$ & ${\small 5}$ & ${\small 35}$\\
${\small (2,2)}$ & ${\small 74}$ & ${\small 7}$ & ${\small 7}$ & ${\small 12}$\\
${\small (4,0)}$ & ${\small 19}$ & ${\small 50}$ & ${\small 4}$ &
${\small 26}$\end{tabular}
}\end{table}To conclude, we plot biased Q-values at a $Qb$-equilibrium
($b^{\ast}=2\varpi$) (see Figure~\ref{fig10}) over long and short lapses of
time, as well as the differences $\Delta_{i}=Q_{i}(C)+b^{*}_{i}%
-Q_{i}(D)$ (see Figure~\ref{F9}). 

In comparison with Figure~\ref{fig6}, which has been drawn over the same long
horizon, Figure~\ref{fig10a} illustrates that under biased Q-learning, players
no longer fall into defection traps under that horizon. Figure~\ref{fig10b}
and~\ref{F9} illustrate that players still alternate between cooperative
phases and disorganized phases, but disorganized phases are shorter and short
enough to avoid a drop in $Q-$values. Joint defections arise, but only within
disorganized phases.

\subsection{Comparative statics.\label{seccomp}}

Under naive Q-learning, players alternate between high-Q phases and low-Q
phases. The role of the bias is to facilitate transitions to (and the
persistence of) high-Q phase. Incentives to raise the bias depends on the
prevalence of low-Q phases. We discuss below comparative statics with respect
to $y$ and $\alpha$.

In order to avoid multiplicity issues, but also as a mean of selecting
equilibria based on a notion of evolutionary stability, we compute for each
matrix obtained the (unique) limit Quantal Response equilibrium (see
\citet{compte23}), obtained by endogenizing the logit parameter as follows:
starting from a logit parameter $\beta$ equal to 0, we gradually increase
$\beta$ until the logit-response dynamic becomes unstable or otherwise we
increase it indefinitely. (See the \hyperref[app3]{Appendix} for a formal
definition). For example, for the gain matrix of Table~\ref{T6a}, the
procedure selects the equilibrium with smallest bias yielding 1.74.

\begin{figure}[h]
\centering
\begin{minipage}{0.45\textwidth}
\centering
\includegraphics[scale=0.5]{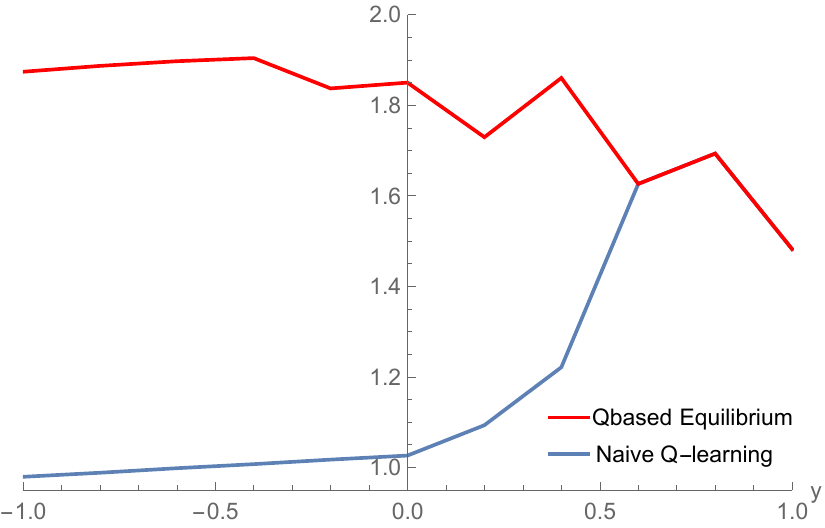}
\subcaption{$\alpha=0.1$}
	\label{figcompa01}
\end{minipage}\hfill\begin{minipage}{0.45\textwidth}
\centering
	 \includegraphics[scale=0.5]{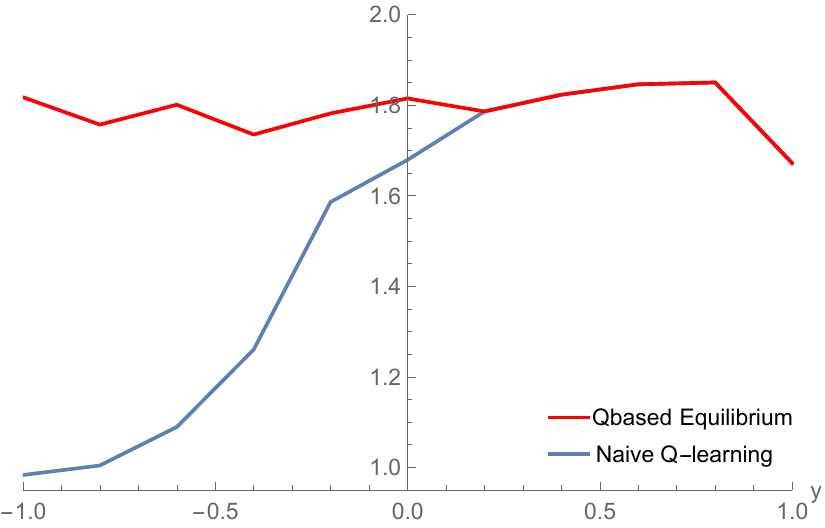}
\subcaption{$\alpha=0.3$}
\label{figcompa01}
\end{minipage}
\caption{Comparative statics. Welfare levels}%
\label{figcomp}%
\end{figure}

We report (see Figure \ref{figcomp}) comparative statics over $y$ for
simulations done over 100000 periods and $\alpha\in\{0.1,0.3\}$, with
$\varepsilon=0.1$, starting from unfavorable conditions. We indicate values
obtained for unbiased and biased Q-learning with biases set at equilibrium values.

For large enough $y$, there is no bias in equilibrium, and equilibrium
outcomes coincide with that obtained under unbiased (naive) Q-learning. The
reason is that when $y$ is closer to $1$, Q-traps are shallower and exits are
easier so high Q phases are preponderant \textit{even when there is no bias}.
In such cases, there is no incentives to distort the policy rule.

As $y$ gets lower, defection traps become an issue, and equilibrium biases
adjust upward, keeping welfare levels almost constant across $y$ and $\alpha$,
so essentially avoiding defection traps.

Figure~\ref{figcomp} also illustrates that, given the horizon and initial
conditions considered, a lower speed of adjustment $\alpha$ increases the
prevalence of low-$Q$ phases even for intermediate values of $y$. In these
cases, biased Q-learning helps.

It is interesting to contrast our observations with \citet{banchio22a} who
find two stable component for the continuous time limit. These stable
components are consistent with our low-Q and high-Q phases. Our simulations,
which are done far from the continuous time limit, allow us to examine
transitions between phases, and how biased Q-learning, in equilibrium, tilts
the outcome in favor of high-Q phases.

\section{Stochastic payoffs\label{sec_stochastic}}

We have so far assumed that payoffs are obtained with certainty. We now assume
that payoffs are \textit{stochastic}, meaning that conditional on the action
profile $a$ played, each player $i$ obtains a random payoff $z_{i}$. In light
of the traditional approach to repeated games, this payoff can be interpreted
as an imperfect \textit{signal} correlated with the other's behavior, so this
assumption moves us to realm of games with \textit{imperfect monitoring}.
Furthermore, if the payoffs are drawn independently (conditional on actions
played), monitoring is \textit{private}.

The distinction between perfect/imperfect and public/private monitoring has
been useful in the traditional approach. With perfect or imperfect public
monitoring, \textit{public information} is available. This allows perfectly
coordinated play where public information is used by players to finely tune
their behavior to others' at any date. Under private monitoring, no such fine
tuning is possible.

With $Q-$based strategies, the distinction between monitoring structures seems
less relevant. Play is only based on one's own payoff experience, i.e. one's
own Q-values. So, even if, in addition to payoff realizations, actions are
perfectly observable, this information about action played is not used to
update Q-values. As a matter of fact, as soon as players experiment and play
differently (either $CD$ or $DC$) -- or if payoffs are asymmetric, players get
\textit{different payoff histories}, and potentially, $\Delta_{1}^{t}$ and
$\Delta_{2}^{t}$ then differ from one another. As we have seen, these
differences are conducive to disorganized phases where play is \textit{not
perfectly coordinated}, with frequent alternation between $CD^{\prime}s$ and
$DD$'s. In these phases, $\Delta_{i}^{t}$ is close to $0$ and even a
sophisticated agent could not predict, conditional on $(Q_{i}^{t},\Delta
_{i}^{t})$, the sign of $\Delta_{j}^{t}$.\footnote{Away from these
disorganized phases however, $\Delta_{i}^{t}$ and $\Delta_{j}^{t}$ have the
same sign, so play is coordinated, as emphasized by \citet{banchio22a}.}


The main insight from Section \ref{section3} is that when defection traps are
an issue (i.e., when $y$ is small enough), players have an incentive to bias
their policy rule, with equilibrium biases fostering high welfare levels.

What happens when payoffs are stochastic? One effect is to increase the
variability of the payoffs obtained, which potentially induces larger
variations in $Q$-values. Stochastic payoffs may also modify the co-evolution
of the pair $(\Delta_{1}^{t},\Delta_{2}^{t})$, in particular if payoff shocks
are independent. We shall see that large variations in $Q$-values may help
coordinated exits from traps (\textit{to the extent that payoff shocks are
correlated), }but they also reduce the durability of cooperative phases. So
being trapped remains an issue, and we shall see that in these cases as well
(i.e., for low $y$) players have incentives to bias their policy rule,
fostering equilibrium welfare levels comparable to those obtained when payoffs
are not stochastic.

\subsection{Payoff structure}

We consider a very simple stochastic payoff structure with only two payoff
realizations $z_{i}\in\{0,V\}$. Cooperating costs $L$, but it increases the
probability of a good outcome. We let $p_{k}$ denote the probability of a good
outcome when $k$ players cooperate, assuming $p_{k}$ increases with $k$. The
expected payoff matrix for player 1 is thus%
\[%
\begin{tabular}
[c]{c|cc}%
$a_{1}\backslash a_{2}$ & $C$ & $D$\\\hline
$C$ & $p_{2}V-L$ & $p_{1}V-L$\\
$D$ & $p_{1}V$ & $p_{0}V$%
\end{tabular}
\ \ \
\]
Any expected payoff matrix examined in the previous Section (parameterized by
$x$ and $y$) can be generated by setting $V>2+x-y$ and then adjusting the
$p_{k}$'s and $L$.\footnote{For a given $x,y$, with $V>2+x-y$, we let
$p_{0}=1/V$, $p_{1}=x/V$, $L=x-y$, and $p_{2}=(2+x-y)/V.$} For the sake of
comparison with the previous Section, we set $x=2.5$, allow $y$ to vary within
the interval $[-1,1]$, and choose $V>5.5$ (thus ensuring that the $p_{k}$'s
are well-defined for all $y\in\lbrack-1,1]$).\footnote{In simulations we set
$V=6$, so $p_{0}=1/6$, $p_{1}=5/12$ and $p_{2}=(9-2y)/12$, so $p_{2}$ varies
between $7/12$ and $11/12$ as $y$ varies from $1$ to $-1$.}

We investigate two cases regarding the correlation between payoff
realizations. We examine the (perfectly) \textbf{correlated case} where
$z_{1}=z_{2}=z$ and the\textbf{ independent case} where the $z_{i}$ are drawn
independently (conditional on the action profile played).

In the correlated case, when $a$ is either $CC$ or $DD$, realized gains are
identical (and equal to $z$ and $z-L$ respectively). When $a=CD$ or $DC$,
payoff realizations are identical, but gains differ: if $CD$ has been played,
then player 1 gets $z-L$ while player 2 gets $z$. In contrast, in the
independent case, payoff realizations may differ even when $CC$ or $DD$ is
played. If $CC$ is played, player $1$ may get a good draw and gain $V-L$ while
player 2 may get a bad draw and gain $-L$.

The main difference with Section~\ref{section3} is that the randomness in
payoff realizations may generate large variations in $Q$-values over a
relatively short time scale, compared to the case with no noise, as sequences
of good draws generate a rise in $Q$-values (plausibly conducive to
cooperation), while sequences of bad draws generate a drop in Q-values,
conducive to defections.

\subsection{The dynamics of Q-values}

\begin{figure}[h]
\centering
\includegraphics[scale=0.5]{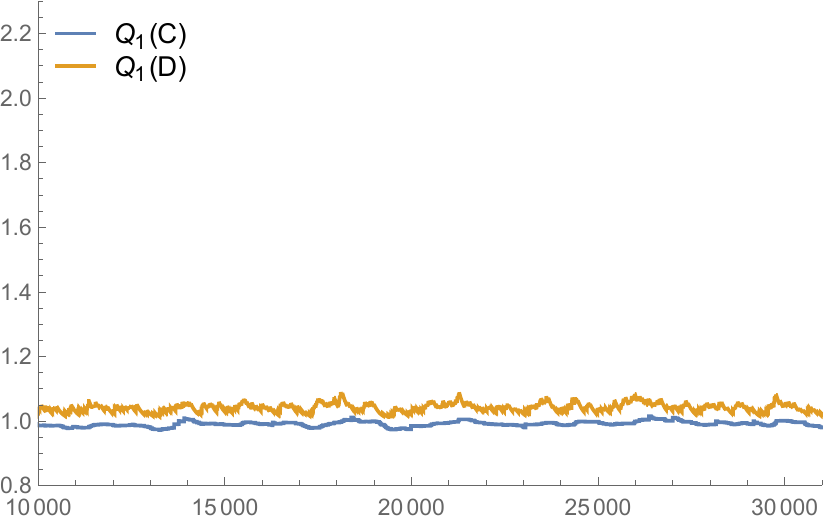} \caption{$Q$-values with no
randomness on payoffs}%
\label{figplotperf}%
\end{figure}

We illustrate the dynamics of Q-values assuming $y=0$, setting $\alpha
=0.1,\varepsilon=0.05$ and $\delta=0.95$ and starting from unfavorable
conditions ($Q_{i}^{0}(D)=1>Q_{i}^{0}(C)=0.9$). In case payoffs are not
random, the low speed of adjustment makes it hard to exit from a defection
trap. Experimentations generate small variations in Q-values, but too small to
generate any exit over the million period draw considered.
Figure~\ref{figplotperf} shows a 20000 period-long realization.

\begin{figure}[h]
\centering
\begin{minipage}{0.45\textwidth}
\centering
\includegraphics[scale=0.5]{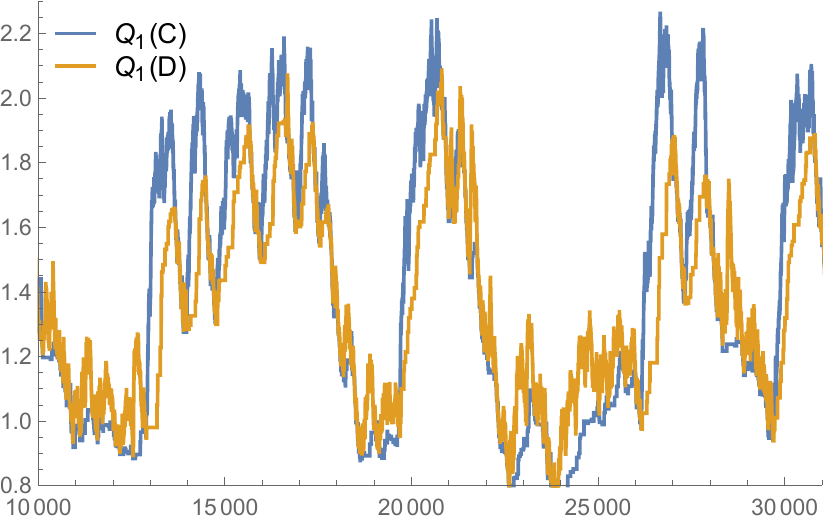}
\subcaption{Correlated shocks}
	\label{figplotcor}
\end{minipage}\hfill\begin{minipage}{0.45\textwidth}
\centering
	 \includegraphics[scale=0.5]{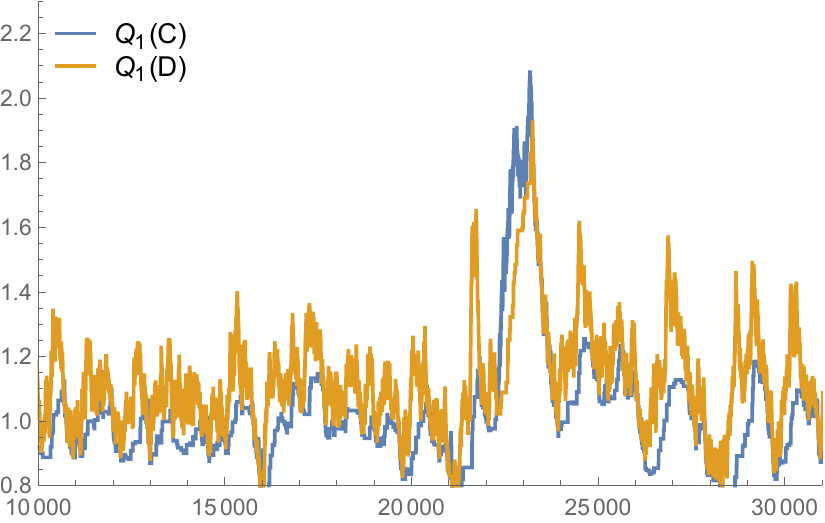}
\subcaption{Independent shocks}
\label{figplotpriv}
\end{minipage}
\caption{The dynamics of Q-values with random payoff shocks}%
\label{figdyn}%
\end{figure}

In case payoffs are random (see Figure \ref{figdyn}), $Q$ values are much more
variable even along defection phases (where $Q_{i}(D)>Q_{i}(C)$), and one
observes transitions between mostly-defective phases and mostly-cooperative
phases whether shocks are correlated or independent.
The Figures suggest however that these transitions are less frequent and less
persistent when payoff shocks are independent. Confirming this, we report
below (Table~{\ref{tab6}) the occurrence of each pair $CC$,$CD,DC$ and $DD$
for $\alpha=0.1$ and various specifications of $\varepsilon$, over draws of
100.000 periods, as well as expected gains. }

\begin{table}[h]
\caption{Gains and frequencies}%
\label{tab6}%
\begin{subtable}{.3\linewidth}
\centering
\caption{no randomness}
\scalebox{0.75}{
\begin{tabular}[c]{c|cccc}
$\varepsilon$ & \small gains & \small \% CC & \small \%CD,DC & \small \%DD \\\hline
0.05 & 1.01 & 0.1 & 2.5 & 95 \\
0.1 & 1.03 & 0.2 & 5 & 90 \\
0.15 & 1.04 & 0.5 & 7 & 86 \\
\end{tabular}
}
\end{subtable}
\hfill\begin{subtable}{.3\linewidth}
\centering
\caption{correlated shocks}
\scalebox{0.75}{
\begin{tabular}[c]{c|cccc}
$\varepsilon$ & \small gains & \small \% CC & \small \%CD,DC & \small \%DD \\\hline
0.05& 1.31 & 27 & 7 & 58 \\
0.1 & 1.19 & 13.5 & 10 & 66 \\
0.15 & 1.13 & 7 & 12.5 & 68 \\
\end{tabular}
}
\end{subtable}
\hfill\begin{subtable}{.3\linewidth}
\centering
\caption{independent shocks}
\scalebox{0.75}{
\begin{tabular}[c]{c|cccc}
$\varepsilon$ & \small gains & \small \% CC & \small \%CD,DC & \small \%DD \\\hline
0.05 &1.06 & 2 & 7  & 84 \\
0.1&1.09 & 4 & 10.5 & 75\\
0.15&1.13 & 7 & 12.5 & 68 \\
\end{tabular}
}
\end{subtable}
\end{table}

\begin{figure}[h]
\centering
\includegraphics[scale=0.5]{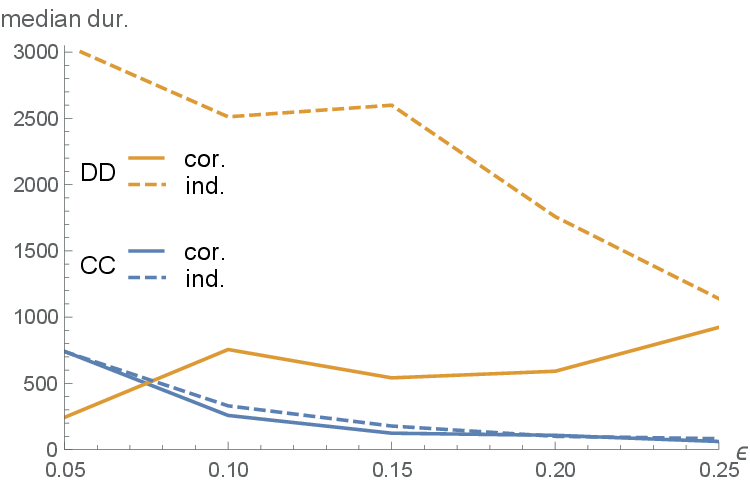}\caption{Median exit durations}%
\label{fig14}%
\end{figure}

The tables confirm that with no randomness, there cannot be joint cooperation
beyond occasional joint experimentation, and that correlated shocks are more
conducive to cooperation phases than independent shocks, though this effect is
reduced when experimentation is more frequent.\footnote{The reason is that
higher experimentation levels shorten significantly cooperative phases without
facilitating much exits from defective phases (see Figure\ \ref{fig14}).}

Intuitively, payoff shocks facilitate exits from defection traps compared the
no-randomness case because they induce large variations in the $Q$-value of
the \textit{currently preferred action}, say $D$. In comparison, $Q_{i}%
^{t}(C)$ varies only occasionally (i.e., when experimented). These differences
in variations of $Q$-values imply that soon enough $Q_{i}^{t}(C)$ will exceed
$Q_{i}^{t}(D)$, with $C$ becoming the preferred action. In comparison, with
non-random payoffs, escaping low-Q traps required\textit{\ joint
experimentation} (hence high experimentation levels)\textit{. }Joint
experimentation is no longer necessary with random shocks.

To give further perspective on the difference between correlated and
independent shocks, we simulate (for each $\varepsilon)$ 25 draws of 10.000
periods and obtain the median duration of a cooperative phase starting from an
exogenously fixed favorable initial condition. We do the same to obtain the
median duration of a defective phase, starting from a fixed unfavorable
initial condition.\footnote{For the favorable conditions we choose
$Q_{i}(C)=1.5$ and $Q_{i}(D)=1.4$ for both players and consider the first date
for which $\Delta_{i}<0$ for two consecutive periods for both players. For the
unfavorable condition we choose $Q_{i}(C)=1.2$ and $Q_{i}(D)=1.25$ and look
for the first date for which $\Delta_{i}>0$ for two consecutive periods for
both players.} We plot these median durations in Figure~\ref{fig14} below.
It reveals that whether shocks are independent or correlated, the
durations of cooperative phase are similar, with more experimentation
shortening the duration of cooperative phases. The main effect of independence
is to substantially raise the length of defection phases, in particular when
experimentation is small.

Intuitively, starting from unfavorable conditions, the time it takes for
\textit{either} $\Delta_{1}$ or $\Delta_{2}$ to rise above $0$ is
\textit{shorter} under independent shocks.
The issue is that independent shocks tend to raise the gap $|\Delta_{i}%
^{t}-\Delta_{j}^{t}|$ even when $DD$ is played and this gap makes it less
likely to subsequently generate a jointly cooperative path: once, say,
$\Delta_{1}^{t}>0>\Delta_{2}^{t}$, $CD$ is played and on average \textit{both}
$\Delta_{i}^{t}$ must decrease (because the cooperative player tends to get
low payoffs -- which lowers $Q_{1}^{t}(C)$, while the defecting player tends
to get good payoffs -- which increases $Q_{2}^{t}(D)$). In contrast, with
correlated shocks, in a defection phase, both $\Delta_{i}^{t}$ either rise
(after a bad draw) or decrease (after a good draw), and these covariations
help a simultaneous rise of $\Delta_{i}^{t}$ above $0$.

\subsection{Q-based learning and equilibrium biases}

We adopt the same methodology as in Section \ref{sectionQbased}. We set
$\alpha=0.1,\delta=0.95$ and $\varepsilon=0.1$ and consider a finite grid of
possible biases $b_{i}\in B_{i}$, where $b_{i}=\kappa_{i}\varpi$ for
$\kappa_{i}\in\{0,1,..,K\}$, for an increment $\varpi=0.02$. For each value of
$y$ considered, we compute the anonymized payoff matrix $\overline{v}%
^{y}(\kappa_{1},\kappa_{2})$ obtained for $K=6$ for the correlated case and
the independent case respectively. We report these matrices in the Appendix,
for different values of $y$.

\begin{figure}[h]
\centering
\begin{minipage}{0.45\textwidth}
\centering
\includegraphics[scale=0.5]{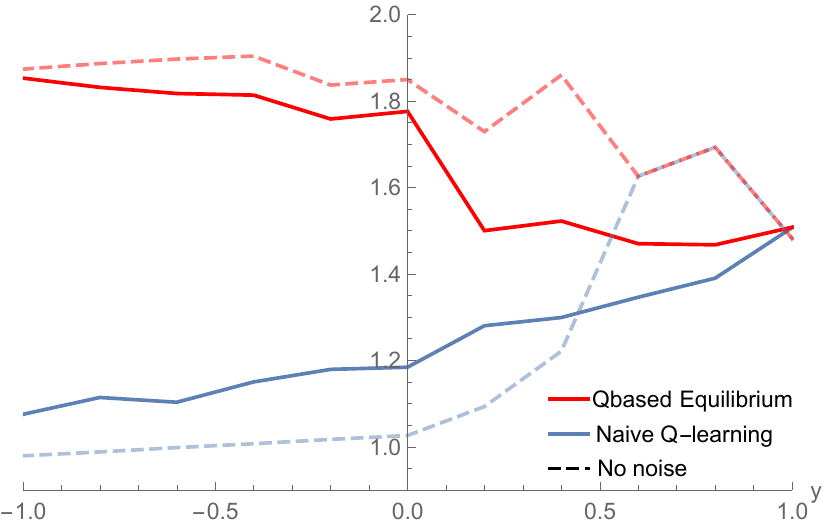}
\subcaption{Correlated shocks}
	\label{figcorcomp}
\end{minipage}\hfill\begin{minipage}{0.45\textwidth}
\centering
	 \includegraphics[scale=0.5]{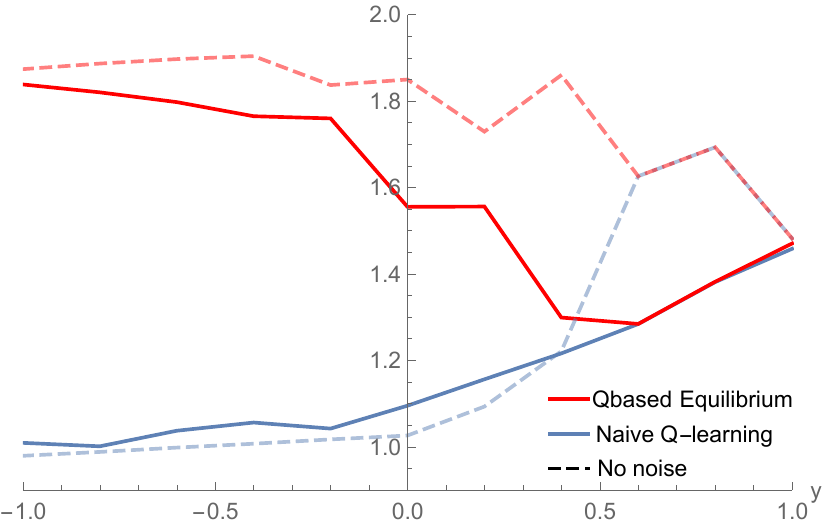}
\subcaption{Independent shocks}
\label{figprivcomp}
\end{minipage}
\caption{Comparative statics}%
\label{figcompcorpriv}%
\end{figure}

From the matrix $\overline{v}^{y}$, the payoff $\overline{v}^{y}(0,0)$ gives
us the expected gain obtained under naive Q-learning. We also use
$\overline{v}^{y}$ to find Q-based equilibria. Because there may be several
equilibria in some cases, we proceed as before and compute the (uniquely
defined) limit Quantal response equilibrium for each matrix $\overline{v}^{y}%
$. Figure~\ref{figcompcorpriv} reports the equilibrium value obtained, as well
as the payoff from naive Q-learning, as we vary $y$. We also indicate how
these payoffs compare to those derived when payoffs are non-stochastic.

Figure~\ref{figcompcorpriv} shows that as $y$ decreases, naive Q-learning
leads to payoffs close to the Nash outcome (1). Correlated shocks help support
some cooperation (above levels obtained under independent shocks or under no
randomness), but the extent of cooperation remains small. In contrast, with
the option to bias Q-learning, the unique Q-based outcome leads to substantial
(and comparable) levels of cooperation, independently of the nature of the
shocks (independent or correlated), for all values of $y$ considered: as $y$
decreases, the equilibrium bias increases and fosters cooperative outcomes,
preventing (as in the previous Section) the occurrence of defection traps.

\section{Duopoly\label{sec_duopoly}}

In this Section, we revisit \citet{calvano20} in light of our approach. We
consider a duopoly game where each player has finitely many price alternatives
$p_{i}\in A_{i}$, with profits $\pi_{i}(p)$ depending on the profile of prices
$p=(p_{i},p_{j})$. Profits are determined by
\[
\pi_{i}(p_{i},p_{j})=10(p_{i}-c)\frac{\exp(\frac{d-p_{i}}{\mu})}{1+\exp
(\frac{d-p_{i}}{\mu})+\exp(\frac{d-p_{j}}{\mu})}%
\]
where $\mu$ is an index of horizontal differentiation (the limit case
$\mu\simeq0$ corresponding to perfect substitutes) and $d$ a demand parameter.
In simulations below, we set $d=2$ and $c=1$, and make comparative statics
with respect to $y\equiv1/\mu$. We also assume a finite number of prices,
$p^{k}=1.4+k0.1$ with $k\in\{0,...,6\}$. To fix ideas, we report the payoff
matrices for $y=6$ and $y=10$. In both cases, the Nash equilibrium obtains for
$k=0$ and the welfare maximizing choice obtains for $k=5$, with the
corresponding payoffs indicated in bold. A higher $y$ results in stronger
incentives to undercut the other player when $k>0$.

\begin{table}[h]
\caption{Payoff matrix examples}%
\label{Tmatrix}
\begin{subtable}{.5\linewidth}
\centering
\caption{$y=6$}
\scalebox{0.8}{
\begin{tabular}{c|ccccccc}
$ k_1\backslash k_2$ & 0 & 1 & 2 & 3 & 4 & 5 &6\\\hline
0&\textbf{1.97} & 2.54 & 3.01 & 3.35 & 3.58 & 3.71 & 3.79 \\
1&1.74 & 2.44 & 3.13 & 3.7 & 4.11 & 4.38 & 4.55 \\
2&1.36 & 2.06 & 2.87 & 3.66 & 4.31 & 4.78 & 5.08 \\
3&0.97 & 1.56 & 2.34 & 3.23 & 4.08 & 4.77 & 5.26 \\
4&0.65 & 1.09 & 1.73 & 2.56 & 3.48 & 4.32 & 4.99 \\
5&0.42 & 0.72 & 1.18 & 1.85 & 2.67 & \textbf{3.53} & 4.29 \\
6&0.26 & 0.45 & 0.77 & 1.24 & 1.88 & 2.62 & 3.33 \\
\end{tabular}
}
\end{subtable} \begin{subtable}{.5\linewidth}
\centering
\caption{$y=10$}
\scalebox{0.8}{
\begin{tabular}{c|ccccccc}
$ k_1\backslash k _2$ & 0 & 1 & 2 & 3 & 4 & 5&6 \\\hline
0 & \textbf{2}. & 2.92 & 3.52 & 3.8 & 3.92 & 3.96 & 3.98 \\
1 & 1.34 & 2.49 & 3.64 & 4.38 & 4.73 & 4.88 & 4.93 \\
2 & 0.71 & 1.61 & 2.97 & 4.33 & 5.2 & 5.62 & 5.79 \\
3 & 0.33 & 0.83 & 1.86 & 3.41 & 4.94 & 5.91 & 6.37 \\
4 & 0.14 & 0.38 & 0.94 & 2.08 & 3.75 & 5.32 & 6.3 \\
5 & 0.06 & 0.16 & 0.42 & 1.03 & 2.2 & \textbf{3.8} & 5.18 \\
6 & 0.02 & 0.07 & 0.18 & 0.45 & 1.06 & 2.12 & 3.33 \\
\end{tabular}
}
\end{subtable}
\end{table}

These games can be thought of generalizations of the prisoner's dilemma to
more than two actions, with $y$ characterizing the strength of competition.


In the spirit of our previous exercise, we study the long-run properties of
naive policy rules where agents take the action with highest $Q$-value (unless
they experiment).\footnote{Note that when players experiment, we assume
uniform experimentation over all available strategies. An alternative
assumption could be that experimentation is only on \textquotedblleft nearby"
strategies, or driven by relative $Q$-values.} We then move to $Qb-$equilibria
where players use biased policy rules. To fix ideas, we use biased policy
rules which take the action that maximizes the biased criteria
\[
Q_{i,b_{i}}(p_{i})=Q_{i}(p_{i})+b_{i}\pi_{i}(p_{i},p_{i})
\]
where $\pi_{i}(p_{i},p_{i})$ is the profit that $i$ would obtain if the other
player was adopting the same price. We analyze the game where each player sets
optimally their own bias $b_{i}$, among a finite number of possible biases,
setting $\alpha=0.1,\delta=0.95$ and $\varepsilon=0.1$. Furthermore, to
compute expected gains, we run $n$ simulations of length $T$, starting from
unfavorable initial conditions,\footnote{$Q_{i}(p^{0})=2$ and $Q_{i}%
(p^{k})=1.8$ for all $k>0$.} and compute the average gain over these $n$
simulations. We set $n=9$ and consider two horizons, $T=100000$ and $T=50000.$
\medskip

\subsection{The dynamics of naive Q-learning}

We first illustrate the dynamics of naive Q-learning assuming $y=6$ and
$T=100000$. We find that players are trapped 60\% of the time in the Nash
profile $(p^{0},p^{0})$, with long-run payoff approximately equal to 2.25.
This payoff is above the Nash level (1.97), partly because experimentation
improves welfare, partly because $(p^{3},p^{3})$ gets some (small) weight
($5\%$). Figure~\ref{fig16} reports the dynamics of $Q_{i}(p^{0})$ and
$\max_{k>0}Q_{i}(p^{k})$ for player $1$: \begin{figure}[h]
\centering
\includegraphics[scale=0.65]{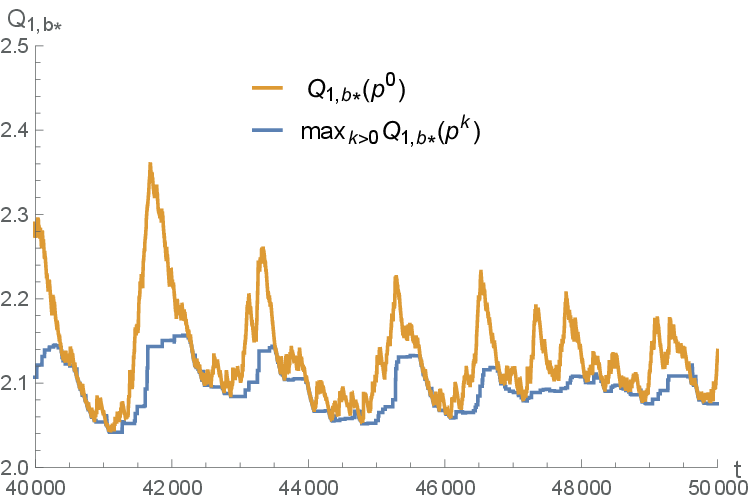}\caption{Evolution of $Q$-values under
naive $Q$-learning}%
\label{fig16}%
\end{figure}

On occasions, player 1 finds that some alternative strategy $p^{k}$ with $k>0$
has higher $Q-$value, but, as it turns out, this is rarely concomitant with
player 2 finding that some $p^{k^{\prime}}$ with $k^{\prime}>0$ has higher
$Q$-value. As a matter of fact, whenever player 1 prefers some $p^{k}>p^{0}$,
this reinforces the attraction of player 2 for $p^{0}$:\ the joint experience
of some pair $(p^{k},p^{k^{\prime}})$ which would yield gains Pareto superior
to Nash gains is not frequent enough.

To illustrate graphically this phenomenon, we plot the differences
\[
\Delta_{i}\equiv Q_{i}(p^{0})-\max_{k>0}Q_{i}(p^{0})
\]
for each player. Over the lapse of time shown, these differences are not
simultaneously positive, and players remain trapped in low prices.
\begin{figure}[h]
\centering
\includegraphics[scale=0.65]{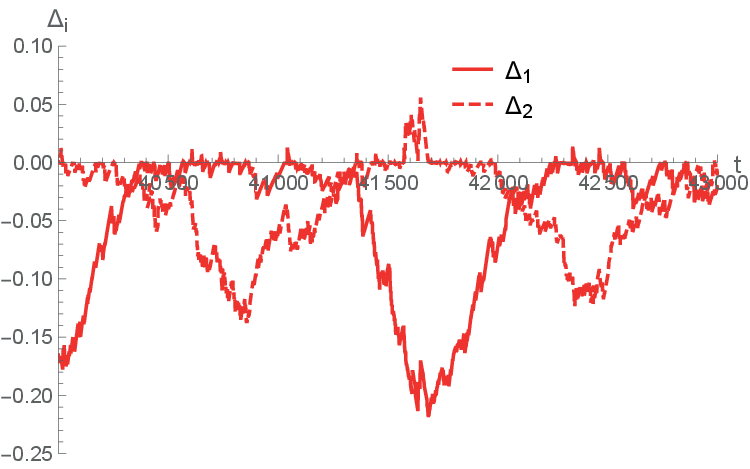}\caption{Co-evolution of $\Delta$
under naive Q-learning}%
\end{figure}

\subsection{Qb-learning and equilibrium biases}

\begin{figure}[h]
\centering
\begin{minipage}{0.45\textwidth}
\centering
\includegraphics[scale=0.53]{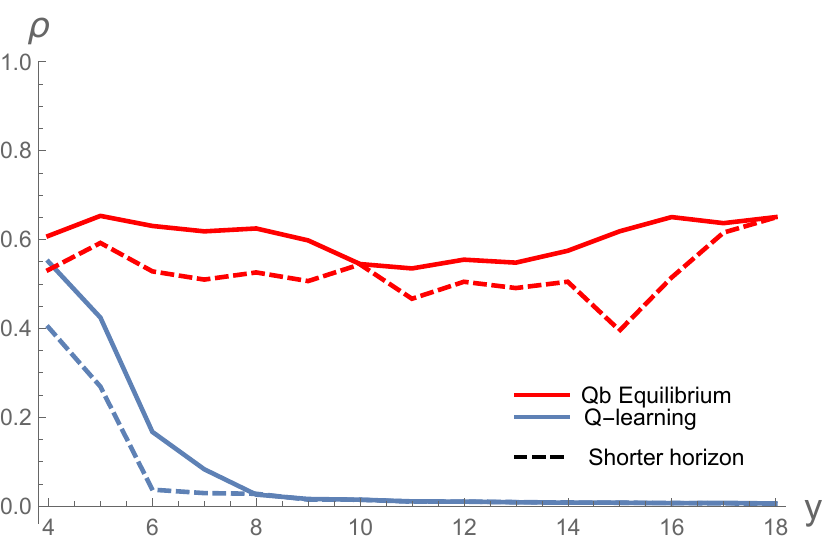}
\subcaption{Collusion measure}
	\label{figcollusion}
\end{minipage}\hfill\begin{minipage}{0.45\textwidth}
\centering
	 \includegraphics[scale=0.5]{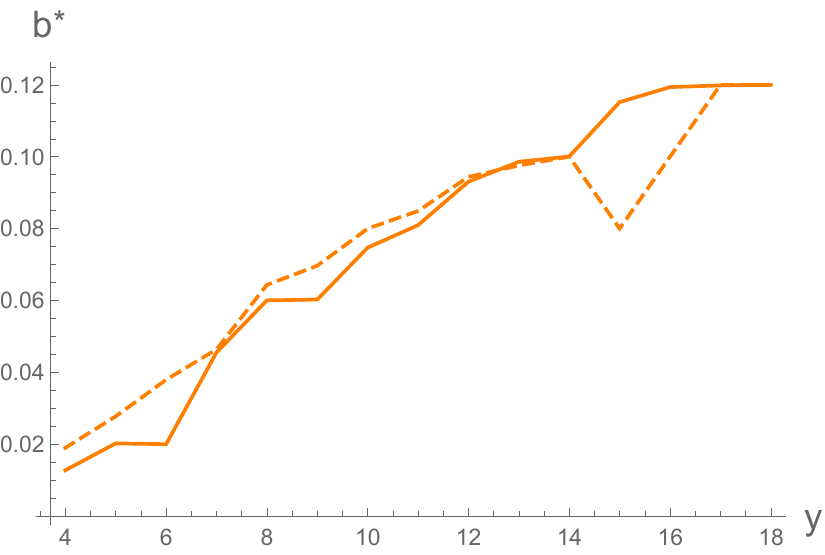}
\subcaption{Equilibrium biases}
\label{figduopolybias}
\end{minipage}
\caption{Duopoly}%
\label{figduopoly}%
\end{figure}

We compute expected gains when players use biased policy rules with
$b_{i}=\kappa_{i}\varpi$ where $\varpi=0.02$, for $\kappa_{i}\in\{0,...,6\}$,
and we do this for $y\in\{4,...,18\}$. As in previous Sections, we derive for
each $y$ a payoff matrix $v^{y}$ which we use to compute equilibrium values
$v_{eq}^{y}$.\footnote{To take care of multiplicity issues, we derive (as
before),\ for each matrix $v^{y}$, the limitQRE equilibrium. The limit QRE may
be mixed, in which care we report the expected bias. Multiplicity arises for
large $y$, and the procedure selects a high-value equilibrium, though not
necessarily the highest value one.} In order to assess the level of collusion
implied, we compare these equilibrium values to the Nash payoff (denoted
$\underline{v}^{y})$ and the largest symmetric gain $\overline{v}^{y}=\max
\pi^{y}(p^{k},p^{k})$. Specifically, we define a measure of collusion as the
fraction
\[
\rho^{y}=(v_{eq}^{y}-\underline{v}^{y})/(\overline{v}^{y}-\underline{v}^{y})
\]
Figure~\ref{figduopoly} reports these levels of collusion for each $y$ and
each horizon $T$ considered.

Some cooperation is possible under naive Q-learning so long as the competition
parameter $y$ is not too large. In contrast, under $Q$-based learning,
equilibrium biases adjust upward as the competition parameter $y$ increases,
inducing significant levels of collusion (around 60\%) and remarkably robust
to $y$. These levels are slightly smaller under the shorter horizon.

\begin{figure}[h]
\centering
\includegraphics[scale=0.65]{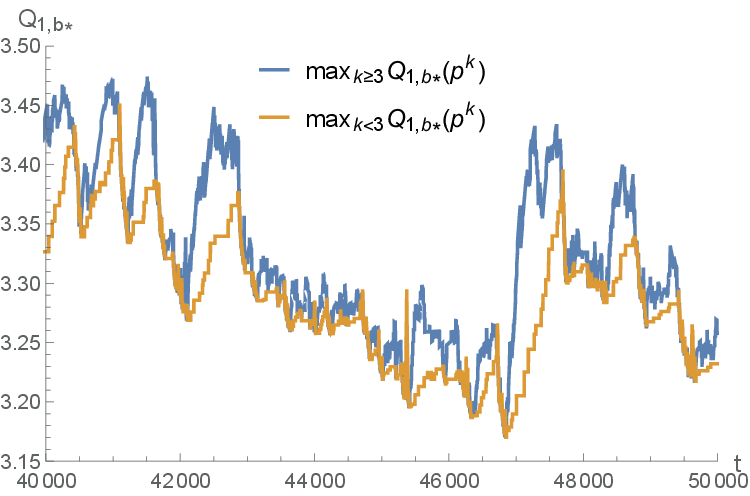}\caption{Evolution of $Q$-values under
$b^{\ast}$}%
\label{F18}%
\end{figure}

We conclude with an illustration of a typical path obtained in a Q-biased
equilibrium, setting $y=6$ (with $b^{\ast}=0.02)$.
Figure~\ref{F18} plots the evolution of $\max_{p<p^{3}}Q_{i,b^{\ast}}^{t}(p)$
and $\max_{p\geq p^{3}}Q_{i,b^{\ast}}^{t}(p)$. 

\begin{figure}[h]
\centering
\includegraphics[scale=0.65]{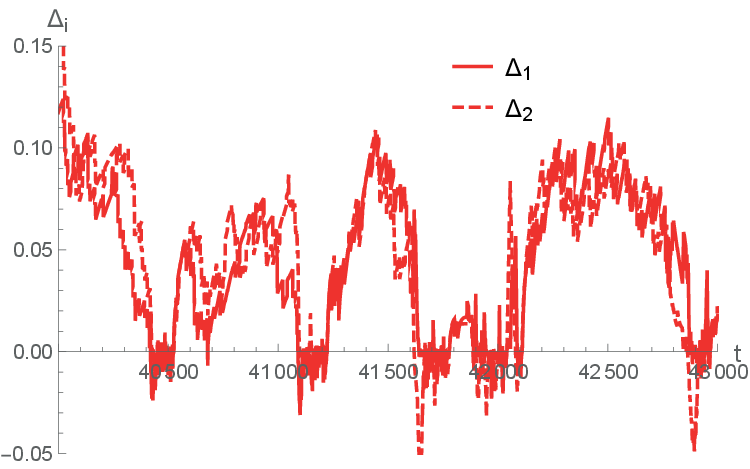}\caption{Co-evolution of $\Delta$
under $b^{*}$}%
\label{fig19}%
\end{figure}

We observe that $Q-$values remain significantly above Nash profits yet below
the maximum joint profits (3.53). Prices mostly remain\ at levels above or
equal to $p^{3}$. To better assess the dynamics when $Q-$values of collusive
and non-collusive prices get close to one-another, we define
\[
\Delta_{i}=\max_{p\geq p^{3}}Q_{i,b^{\ast}}(p)-\max_{p<p^{3}}Q_{i,b^{\ast}%
}(p).
\]
When $\Delta_{i}$ is positive, player $i$ plays a price above or equal $p^{3}%
$, and when $\Delta_{i}$ is negative, player $i$ plays a price equal to
$p^{0},p^{1}$ or $p^{2}$. Figure \ref{fig19} plots the co-evolution of
$\Delta_{1}$ and $\Delta_{2}$. 

This co-evolution is analogous to that obtained for the prisoner's dilemma
when $Q$-values are high: there are collusive phases when both $\Delta$'s are
high and these phases contribute to a simultaneous surge of $\max Q$. But
there are episodic \textit{disorganized phases} where the $\Delta$'s remain
close to $0$ and players alternate between prices above and below $p^{3}$. At
the $Qb$ equilibrium however, these disorganized phases are not long enough to
trigger price wars where both would durably play $p<p^{3}$, while in the
absence of biases, these disorganized phases would lead to a long price war
where $p^{0}$ would be preponderant.

\section{Conclusion\label{sec_conclusion}}

Collusive or cooperative behavior requires some form of history-dependent
behavior, whereby the use of competitive strategies by one player are
punished, for example by the use of more competitive strategies by others, to
some extent. A natural motivation for history-dependent behavior is
\textit{learning}. If there is some \textit{uncertainty} about what others do,
and some persistence in what others do, then it may be worth using past data
to better adapt one's behavior to others'.

The source of the uncertainty may be some exogenous (and persistent) shifts in
the value of cooperation, as one then has to identify \textit{when}
cooperation is worthy. The source of uncertainty may also be endogenous,
because it can be that others find, given their observations, and possibly
erroneously, that cooperation is not reciprocated hence being cooperative is
not worthy.

This (uncertainty-based) \textit{learning} motivation for history-dependent
strategies has rarely been the focus of the repeated game
literature,\footnote{This is unlike the reputation literature of course, where
learning is central (\citet{fudenberg89}).} mostly for technical reasons:
finding optimal strategies is easier when there is no exogenous and persistent
uncertainty that affects the payoff structure; equilibria are also simpler to
design when the signals about payoffs are public and one focuses on public
strategies. History-dependence is obtained in equilibrium, but mostly through
a well-designed coordination of continuation strategies, rather than as a
by-product of individual learning from past data in an uncertain
environment.\footnote{With private monitoring, there has been attempts to
construct (belief-based) equilibria where players effectively use signals to
learn about other's past behavior (\cite{compte94,compte02} and
\citet{sekiguchi97}). However, most of the literature has followed the
belief-free path set by (\citet{piccione02} and \citet{ely02}) to construct
finely designed strategies where at coordinated times, most of past history
can be ignored (see also \citet{sugaya22}). These papers rely on player'
sharing a common clock to coordinate play and adjust incentives.}

The algorithmic perspective departs from the classic approach to repeated
games. It incorporates\textit{ }some exogenous uncertainty (assuming an
exogenous probability of experimentation at any stage), and it precludes the
use of perfectly coordinated strategies (assuming that behavior may only be
conditioned on private statistics about one's own past payoffs). These
assumptions justify that players would wish to learn from past experience, but
the characterization of optimal strategies (i.e., Bayesian learning) is not
addressed. The approach assumes a specific learning rule, such as Q-learning.

In other words, the classic path focuses on incentives and essentially leaves
no role to learning, while the algorithmic path puts (non-Bayesian) learning
at the forefront, but without checking incentives, fixing exogenously how past
experience is used.

In this paper, we put learning at the
forefront, through the use of Q-based algorithms, but we do not take for
granted that players would stick to using a suboptimal algorithm (such as
naive Q-learning). We allow each player to \textquotedblleft learn" which
algorithm (in a given class) performs best for her. For one, this added
sophistication allows players to choose a strategy that mostly ignores past
private data, so we do \textit{not} assume history-dependence: we effectively
check that cooperation is not just a by-product of players being constrained
to use a suboptimal strategy (i.e., naive Q-learning). Second, it permits us
to see that with added sophistication, cooperation more easily prevails,
independently of initial Q-values, and even in environments where payoff
signals are stochastic and drawn from conditionally independent distributions,
contrasting with the case where only naive Q-learning would be considered.
Last, by examining the induced game over algorithms and the logit-response
dynamics, we find that lenient algorithms enabling significant levels of
cooperation or collusion are selected, justifying our view that collusion is
easily learned.

These findings make the classic distinction between perfect, imperfect public
and imperfect private monitoring, inherited from the repeated game literature,
less relevant than generally thought. From a practical perspective, the large
scope for collusion obtained is a clear challenge to regulatory bodies.
Players need not observe or condition behavior on others to sustain it.
Lenient behavioral policies do not require any pre-play agreement but easily
emerge from best or logit-response dynamics. Once players use such policies,
collusive outcomes arise spontaneously independently of initial conditions,
whether payoffs are subject to shocks or not, and for a broad range of payoff
conditions: more competitive environments just give rise to more lenient policies.

Finally, as a question for further research, one might investigate whether
similar conclusions hold for other reinforcement learning rules such as UCB
(where each arm is evaluated independently) or when the family of automata
considered is enriched (even if in practice there is obviously a limit to the
number of automata that a player can correctly evaluate). We conjecture that,
as for the repeated game literature, a richer set would just enlarge the scope
for collusion, without undermining it. As an illustration supportive of that
conjecture, we report in the Appendix an example where players can choose both
the bias $b_{i}$ and the speed of adjustment $\alpha_{i}$.


\bibliographystyle{chicago}
\bibliography{reflib}

\newpage

\section*{Appendix}

\subsubsection*{\label{app0}Limit Quantal Response Equilibria}

Consider a game characterized by a set $A$ of action profiles and payoff
vectors $v=(v(a))_{a\in A}$. For any $\lambda\geq0$, and any $\alpha
=(\alpha_{i})_{i}\in\times_{i}\Delta(A_{i})$, define the logit response
$\alpha^{\prime}=h_{\lambda}(\alpha)$ where for player $i,$ $\alpha
_{i}^{\prime}$ is proportional to $\exp\lambda v_{i}(a_{i},\alpha_{-i})$. By
induction on $n$, we define $h_{\lambda}^{(n)}(\alpha)\equiv h_{\lambda
}(h_{\lambda}^{(n-1)}(\alpha))$, and write $h_{\lambda}^{\infty}(\alpha) $
when the limit as $n$ rises converges. By construction $h_{\lambda}^{\infty
}(\alpha)$ is a quantal response equilibrium.

We choose a small increment $\nu$ and set $\lambda_{k}=k\nu$ for
$k\in\mathcal{N}$. At $\lambda_{0}=0$, $h_{0}^{\infty}$ is well-defined and
induces the uniform distribution over actions for each player, which we denote
$\alpha^{0}$. Then, by induction on $k\geq1,$ and so long as $h_{\lambda_{k}%
}^{\infty}(\alpha_{k-1})$ is well-defined, we define the sequence $\alpha
^{k}=h_{\lambda_{k}}^{\infty}(\alpha^{k-1})$. The procedure selects a
(possibly infinite) limit precision $\lambda^{\ast}$ and a Quantal Response
Equilibrium $\alpha^{\ast}$, which we refer to as a limit QRE.

\subsubsection*{\label{app2}Prisoners' dilemma. Perfect information.}

We report here payoff matrices where we check for biases that include negative
values, i.e. $b_{i}=\kappa_{i}\varpi$ with $\kappa_{i}\in\{-1,0,..,4\}$ and
$\varpi=0.02$. To compute these payoffs, we perform $n=8$ simulations over the
horizon $T=100000$, starting for unfavorable conditions. We set $y=-0.5$,
$\varepsilon=0.1$ and compare two values of $\alpha$

\begin{table}[h]
\caption{gain matrices, $y=-0.5$}%
\label{T13}%
\centering
\begin{subtable}{.5\linewidth}
\centering
\caption{$\alpha=0.5$}
\scalebox{0.85}{
\begin{tabular}{c|cccccc}
$\kappa _1\backslash \kappa _2$ & -1 & 0 & 1 & 2 & 3 & 4 \\\hline
-1 & 1.01 & 1.02 & 1.04 & 1.1 & 1.18 & 2.02 \\
0 & 1.02 & 1.2 & 1.52 & 1.69 & 1.74 & 1.89 \\
1 & 1.02 & 1.48 &\textbf{1.68} & 1.76 & 1.8 & 1.83 \\
2 & 1.04 & 1.55 & 1.67 &\textbf{1.77} & 1.83 & 1.87 \\
3 & 1.02 & 1.46 & 1.6 & 1.71 &1.81 & 1.88 \\
4 & 0.13 & 0.49 & 1.23 & 1.52 & 1.68 & 1.87 \\
\end{tabular}
}
\end{subtable}\hfill\begin{subtable}{.5\linewidth}
\centering
\caption{$\alpha=0.1$}
\scalebox{0.85}{
\begin{tabular}{c|cccccc}
$\kappa _1\backslash \kappa _2$& -1 & 0 & 1 & 2 & 3 & 4 \\\hline
-1 & 1.002 & 1.002 & 1.002 & 1.003 & 1.024 & 2.002 \\
0 & 1.002 &\textbf{1.002} & 1.002 & 1.003 & 1.024 & 1.891 \\
1 & 1.002 & 1.002 & 1.002 & 1.034 & 1.21 & 1.871 \\
2 & 1.002 & 1.002 & 1.028 & 1.622 & 1.791 & 1.909 \\
3 & 0.982 & 0.982 & 1.117 & 1.75 &\textbf{1.901} & 1.903 \\
4 & 0.135 & 0.381 & 0.961 & 1.786 & 1.9 & 1.902 \\
\end{tabular}
}
\end{subtable}
\end{table}

When $\alpha=0.5$, there are strong individual up-ward pressures away from
$\kappa=(0,0)$: the gain increases from 1.2 to 1.55 by individually setting
$\kappa_{i}=2$. When $\alpha=0.1$, there are no such individual pressures
because over the horizon considered, exit from the defection trap is too hard.
The consequence is that $\kappa=(0,0)$ is a Qb-equilibrium. Nevertheless this
equilibrium is not stable: our limit QRE selects the Pareto superior one in
this case, in line with risk dominance.

Table~\ref{Tperfect}-\ref{Tperfect03} gather the payoff matrices obtained as
$y$ varies for $\alpha=0.1$ and $\alpha=0.3$.

\begin{table}[h]
\caption{Gain matrices, $\alpha=0.1$}%
\label{Tperfect}
\begin{subtable}{.33\linewidth}
\centering
\caption{$y=-0.8$}
\scalebox{0.65}{
\begin{tabular}{c|ccccc}
$ \kappa _1\backslash \kappa _2$ & 0 & 1 & 2 & 3 & 4 \\\hline
0 & 0.988 & 0.988 & 0.988 & 0.991 & 1.081 \\
1 & 0.988 & 0.988 & 0.988 & 0.991 & 1.445 \\
2 & 0.988 & 0.988 & 1.231 & 1.308 & 1.789 \\
3 & 0.985 & 0.985 & 1.289 & \cellcolor{gray!25}1.885 & \cellcolor{gray!25}1.889 \\
4 & 0.939 & 1.155 & 1.692 & \cellcolor{gray!25}1.884 &\cellcolor{gray!25} 1.889 \\
\end{tabular}
}
\end{subtable}\hfill\begin{subtable}{.33\linewidth}
\centering
\caption{$y=-0.6$}
\scalebox{0.65}{
\begin{tabular}{c|ccccc}
$ \kappa _1\backslash \kappa _2$ & 0 & 1 & 2 & 3 & 4 \\\hline
0 & 0.998 & 0.998 & 0.998 & 1.009 & 1.551 \\
1 & 0.998 & 0.998 & 1.039 & 1.035 & 1.866 \\
2 & 0.997 & 1.033 & 1.503 & 1.693 & 1.895 \\
3 & 0.987 & 1.006 & 1.655 &\cellcolor{gray!25} 1.896 & \cellcolor{gray!25}1.898 \\
4 & 0.792 & 1.163 & 1.779 & \cellcolor{gray!25}1.896 & \cellcolor{gray!25}1.898 \\
\end{tabular}
}
\end{subtable}\hfill\begin{subtable}{.33\linewidth}
\centering
\caption{$y=-0.4$}
\scalebox{0.65}{
\begin{tabular}{c|ccccc}
$ \kappa _1\backslash \kappa _2$ & 0 & 1 & 2 & 3 & 4 \\\hline
0 & 1.007 & 1.007 & 1.009 & 1.139 & 2.315 \\
1 & 1.007 & 1.066 & 1.066 & 1.684 & 2.086 \\
2 & 1.006 & 1.056 & 1.778 & 1.864 & 1.922 \\
3 & 1.005 & 1.422 & 1.823 &\cellcolor{gray!25} 1.907 & \cellcolor{gray!25}1.908 \\
4 & -0.094 & 0.392 & 1.803 &\cellcolor{gray!25}1.907 & \cellcolor{gray!25}1.908 \\
\end{tabular}
}
\end{subtable}
\par
\medskip
\par
\begin{subtable}{.33\linewidth}
\centering
\caption{$y=-0.2$}
\scalebox{0.65}{
\begin{tabular}{c|ccccc}
$ \kappa _1\backslash \kappa _2$ & 0 & 1 & 2 & 3 & 4 \\\hline
0 & 1.017 & 1.017 & 1.025 & 1.604 & 2.397 \\
1 & 1.017 & 1.123 & 1.383 & 1.905 & 2.365 \\
2 & 1.011 & 1.319 &\cellcolor{gray!25} 1.847 &\cellcolor{gray!25} 1.902 & 1.933 \\
3 & 1.013 & 1.437 & \cellcolor{gray!25}1.852 & \cellcolor{gray!25}1.917 & 1.917 \\
4 & -0.024 & 0.102 & 1.828 & 1.917 & 1.917 \\
\end{tabular}
}
\end{subtable}\hfill\begin{subtable}{.33\linewidth}
\centering
\caption{$y=0$}
\scalebox{0.65}{
\begin{tabular}{c|ccccc}
$ \kappa _1\backslash \kappa _2$ & 0 & 1 & 2 & 3 & 4 \\\hline
0 & 1.026 & 1.027 & 1.334 & 2.285 & 2.397 \\
1 & 1.026 & 1.401 &\cellcolor{gray!10} 1.825 & 1.994 & 2.365 \\
2 & 1.155 &\cellcolor{gray!10} 1.692 &\cellcolor{gray!35} 1.874 &\cellcolor{gray!10} 1.935 & 1.942 \\
3 & 0.288 & 1.148 & \cellcolor{gray!10}1.879 & 1.927 & 1.927 \\
4 & 0.155 & 0.278 & 1.854 & 1.927 & 1.927 \\
\end{tabular}
}
\end{subtable}\hfill\begin{subtable}{.33\linewidth}
\centering
\caption{$y=0.2$}
\scalebox{0.65}{
\begin{tabular}{c|ccccc}
$ \kappa _1\backslash \kappa _2$ & 0 & 1 & 2 & 3 & 4 \\\hline
0 & 1.093 & 1.248 & 1.621 & 2.398 & 2.398 \\
1 & 1.183 &\cellcolor{gray!25}\textbf{1.729} & 1.917 & 2.363 & 2.363 \\
2 & 1.202 & 1.703 & 1.888 & 1.944 & 1.949 \\
3 & 0.335 & 0.456 & 1.899 & 1.936 & 1.936 \\
4 & 0.335 & 0.456 & 1.881 & 1.936 & 1.936 \\
\end{tabular}
}
\end{subtable}
\par
\medskip
\par
\begin{subtable}{.33\linewidth}
\centering
\caption{$y=0.4$}
\scalebox{0.65}{
\begin{tabular}{c|ccccc}
$ \kappa _1\backslash \kappa _2$ & 0 & 1 & 2 & 3 & 4 \\\hline
0 & 1.221 & 1.712 & 2.223 & 2.398 & 2.398 \\
1 & 1.497 & \cellcolor{gray!25}\textbf{1.86} & 1.968 & 2.362 & 2.362 \\
2 & 0.675 & 1.649 & 1.933 & 1.953 & 1.957 \\
3 & 0.516 & 0.63 & 1.919 & 1.946 & 1.946 \\
4 & 0.516 & 0.63 & 1.907 & 1.946 & 1.946 \\
\end{tabular}
}
\end{subtable}\hfill\begin{subtable}{.33\linewidth}
\centering
\caption{$y=0.6$}
\scalebox{0.65}{
\begin{tabular}{c|ccccc}
$ \kappa _1\backslash \kappa _2$ & 0 & 1 & 2 & 3 & 4 \\\hline
0 & \cellcolor{gray!25}\textbf{1.626} & 1.612 & 2.395 & 2.399 & 2.399 \\
1 & 1.348 & 1.881 & 2.325 & 2.36 & 2.36 \\
2 & 0.696 & 0.865 & 1.948 & 1.961 & 1.963 \\
3 & 0.696 & 0.806 & 1.936 & 1.955 & 1.955 \\
4 & 0.696 & 0.806 & 1.93 & 1.955 & 1.955 \\
\end{tabular}
}
\end{subtable}\hfill\begin{subtable}{.33\linewidth}
\centering
\caption{$y=0.8$}
\scalebox{0.65}{
\begin{tabular}{c|ccccc}
$ \kappa _1\backslash \kappa _2$ & 0 & 1 & 2 & 3 & 4 \\\hline
0 & \cellcolor{gray!25}\textbf{1.693} & 2.144 & 2.397 & 2.399 & 2.399 \\
1 & 1.038 & 1.659 & 2.356 & 2.358 & 2.358 \\
2 & 0.876 & 0.978 & 1.959 & 1.97 & 1.971 \\
3 & 0.876 & 0.978 & 1.951 & 1.964 & 1.964 \\
4 & 0.876 & 0.978 & 1.948 & 1.964 & 1.964 \\
\end{tabular}
}
\end{subtable}
\end{table}

\begin{table}[h]
\caption{Gain matrices, $\alpha=0.3$}%
\label{Tperfect03}
\begin{subtable}{.33\linewidth}
\centering
\caption{$y=-0.8$}
\scalebox{0.65}{
\begin{tabular}{c|ccccc}
$ \kappa _1\backslash \kappa _2$ & 0 & 1 & 2 & 3 & 4 \\\hline
0 & 1.004 & 1.064 & 1.151 & 1.354 & 1.479 \\
1 & 1.057 & 1.445 & 1.716 & 1.759 & 1.795 \\
2 & 1.115 & 1.635 & \cellcolor{gray!25}\textbf{1.757} & 1.815 & 1.846 \\
3 & 1.206 & 1.553 & 1.715 & 1.839 & 1.877 \\
4 & 1.092 & 1.378 & 1.604 & 1.822 & 1.871 \\
\end{tabular}
}
\end{subtable}\hfill\begin{subtable}{.33\linewidth}
\centering
\caption{$y=-0.6$}
\scalebox{0.65}{
\begin{tabular}{c|ccccc}
$ \kappa _1\backslash \kappa _2$ & 0 & 1 & 2 & 3 & 4 \\\hline
0 & 1.089 & 1.266 & 1.563 & 1.681 & 1.591 \\
1 & 1.243 & 1.624 & 1.762 & 1.809 & 1.811 \\
2 & 1.438 & 1.662 &  \cellcolor{gray!25}\textbf{1.801} & 1.85 & 1.891 \\
3 & 1.382 & 1.571 & 1.751 & 1.855 & 1.894 \\
4 & 0.866 & 1.167 & 1.549 & 1.84 & 1.886 \\
\end{tabular}
}
\end{subtable}\hfill\begin{subtable}{.33\linewidth}
\centering
\caption{$y=-0.4$}
\scalebox{0.65}{
\begin{tabular}{c|ccccc}
$ \kappa _1\backslash \kappa _2$ & 0 & 1 & 2 & 3 & 4 \\\hline
0 & 1.26 & 1.623 & 1.722 & 1.711 & 2.115 \\
1 & 1.568 &  \cellcolor{gray!25}\textbf{1.735} & 1.806 & 1.841 & 1.965 \\
2 & 1.539 & 1.692 & 1.832 & 1.874 & 1.911 \\
3 & 1.33 & 1.548 & 1.776 & 1.877 & 1.906 \\
4 & 0.166 & 0.663 & 1.606 & 1.861 & 1.901 \\
\end{tabular}
}
\end{subtable}
\par
\medskip
\par
\begin{subtable}{.33\linewidth}
\centering
\caption{$y=-0.2$}
\scalebox{0.65}{
\begin{tabular}{c|ccccc}
$ \kappa _1\backslash \kappa _2$ & 0 & 1 & 2 & 3 & 4 \\\hline
0 & 1.586 & 1.754 & 1.789 & 1.673 & 2.307 \\
1 & 1.673 &  \cellcolor{gray!25}\textbf{1.782} & 1.841 & 1.87 & 2.25 \\
2 & 1.564 & 1.706 & 1.852 & 1.909 & 1.94 \\
3 & 1.098 & 1.409 & 1.788 & 1.897 & 1.917 \\
4 & 0.103 & 0.367 & 1.582 & 1.882 & 1.914 \\
\end{tabular}
}
\end{subtable}\hfill\begin{subtable}{.33\linewidth}
\centering
\caption{$y=0$}
\scalebox{0.65}{
\begin{tabular}{c|ccccc}
$ \kappa _1\backslash \kappa _2$ & 0 & 1 & 2 & 3 & 4 \\\hline
0 & 1.679 & 1.794 & 1.827 & 2.069 & 2.388 \\
1 & 1.694 &  \cellcolor{gray!25}\textbf{1.816} & 1.869 & 1.922 & 2.328 \\
2 & 1.538 & 1.707 & 1.881 & 1.924 & 2.044 \\
3 & 0.541 & 1.138 & 1.817 & 1.913 & 1.928 \\
4 & 0.187 & 0.398 & 1.293 & 1.902 & 1.925 \\
\end{tabular}
}
\end{subtable}\hfill\begin{subtable}{.33\linewidth}
\centering
\caption{$y=0.2$}
\scalebox{0.65}{
\begin{tabular}{c|ccccc}
$ \kappa _1\backslash \kappa _2$ & 0 & 1 & 2 & 3 & 4 \\\hline
0 &  \cellcolor{gray!25}\textbf{1.786} & 1.826 & 1.762 & 2.282 & 2.389 \\
1 & 1.701 & 1.842 & 1.905 & 2.179 & 2.334 \\
2 & 1.343 & 1.621 & 1.901 & 1.938 & 2.178 \\
3 & 0.471 & 0.78 & 1.85 & 1.927 & 1.938 \\
4 & 0.364 & 0.555 & 1.058 & 1.918 & 1.936 \\
\end{tabular}
}
\end{subtable}
\par
\medskip
\par
\begin{subtable}{.33\linewidth}
\centering
\caption{$y=0.4$}
\scalebox{0.65}{
\begin{tabular}{c|ccccc}
$ \kappa _1\backslash \kappa _2$ & 0 & 1 & 2 & 3 & 4 \\\hline
0 &  \cellcolor{gray!25}\textbf{1.823} & 1.866 & 1.998 & 2.388 & 2.39 \\
1 & 1.698 & 1.865 & 1.931 & 2.318 & 2.333 \\
2 & 0.946 & 1.574 & 1.919 & 1.952 & 2.192 \\
3 & 0.543 & 0.741 & 1.875 & 1.941 & 1.947 \\
4 & 0.541 & 0.72 & 1.164 & 1.935 & 1.946 \\
\end{tabular}
}
\end{subtable}\hfill\begin{subtable}{.33\linewidth}
\centering
\caption{$y=0.6$}
\scalebox{0.65}{
\begin{tabular}{c|ccccc}
$ \kappa _1\backslash \kappa _2$ & 0 & 1 & 2 & 3 & 4 \\\hline
0 &  \cellcolor{gray!25}\textbf{1.846} & 1.809 & 2.236 & 2.39 & 2.391 \\
1 & 1.541 & 1.891 & 2.054 & 2.331 & 2.332 \\
2 & 0.849 & 1.248 & 1.935 & 1.963 & 2.166 \\
3 & 0.719 & 0.886 & 1.901 & 1.953 & 1.956 \\
4 & 0.718 & 0.886 & 1.356 & 1.952 & 1.955 \\
\end{tabular}
}
\end{subtable}\hfill\begin{subtable}{.33\linewidth}
\centering
\caption{$y=0.8$}
\scalebox{0.65}{
\begin{tabular}{c|ccccc}
$ \kappa _1\backslash \kappa _2$ & 0 & 1 & 2 & 3 & 4 \\\hline
0 & \cellcolor{gray!25}\textbf{ 1.85} & 1.928 & 2.388 & 2.391 & 2.391 \\
1 & 1.312 & 1.919 & 2.271 & 2.329 & 2.329 \\
2 & 0.898 & 1.118 & 1.952 & 1.973 & 2.074 \\
3 & 0.895 & 1.052 & 1.926 & 1.964 & 1.965 \\
4 & 0.895 & 1.052 & 1.691 & 1.964 & 1.964 \\
\end{tabular}
}
\end{subtable}
\end{table}

The gray cells indicates the distribution over biases at the limitQRE. When
there is a single cell, the limitQRE is a Nash equilibrium selection. When
$\alpha=0.3$, all outcomes are pure Nash equilibria. These Nash outcomes are
typically not the Pareto superior ones, but they all involve very substantial
levels of cooperation.

When $\alpha=0.1$, $\kappa=(0,0)$ is typically an equilibrium when $y$ is low,
but a risk dominated one, and it is not a limit QRE. When $y=0$, there is no
pure equilibrium, so the logit-response dynamics becomes unstable for large
enough $\lambda$, and the limitQRE selects an outcome that puts most of the
weight on $\kappa=(2,2)$. $y=-0.2$ illustrates an interesting case where
$\kappa=(3,3)$ is an equilibrium and yet the limit QRE fails to converge to
it. The reason is that when $\kappa_{i}=3$ gets some positive weight,
$\kappa_{i}=4$ must get positive weight as well, which fuels lower level of
$\kappa_{i}$. Nevertheless, most weight is put on $\kappa_{i}\geq2$.

A pure strategy equilibrium may fail to exist. This failure can be a
by-product of our discretization of the space of biases (alternative
discretizations may be conducive to pure equilibria instead). It can also be a
by-product of the fact that simulations do not give us exact expected payoffs,
but only approximations of these expected payoffs. This inherent estimation
noise also justifies that we directly allow for stochastic choice in our predictions.

Regarding the mixed strategy prediction obtained, we interpret it as a
population equilibrium where each player would be involve in many
interactions, and choose a bias that works well across interactions.

\newpage

\subsubsection*{\label{rich}A richer set of automata.}

I assume below that players can choose both the speed of adjustment
$\alpha_{i}\in\{0.1,0.3,0.5\}$ and the bias $b_{i}=\kappa_{i}\omega$ with
$\kappa_{i}\in\{0,1,2,3\}$ and $\omega=0.02$. Each player thus has 12 automata
available characterized by a pair $a_{i}=(\alpha_{i},\kappa_{i})$, indexed as
described in Table~\ref{Tstrategy}. \begin{table}[h]
\caption{Indexing the strategy space}%
\label{Tstrategy}%
\centering
\centering
\vspace{0.3cm} \scalebox{0.8}{
\begin{tabular}{c|cccccccccccc}
\text{$a_i $} & 1 & 2 & 3 & 4 & 5 & 6 & 7 & 8 &9&10&11&12 \\\hline
\text{$(\alpha_i,\kappa_i )$} & (0.1,0) &(0.1,1) &(0.1,2) & (0.1,3) &  (0.3,0) &(0.3,1) &(0.3,2) & (0.3,3)& (0.5,0) &(0.5,1) &(0.5,2) & (0.5,3) \\
\end{tabular}}\end{table}

Tables~\ref{Tlong} and \ref{Tshort} report the gain matrices obtained by
averaging 10 simulations made over a given horizon, for horizons $T=10000$ and
$T=100000$.

\begin{table}[h]
\caption{Gain matrix. Long horizon ($T=100000$)}%
\label{Tlong}%
\centering
\centering
\vspace{0.3cm} \scalebox{0.7}{
\begin{tabular}{c|cccccccccccc}
\text{$a_1\backslash a_2 $} & 1 & 2 & 3 & 4 & 5 & 6 & 7 & 8 &9&10&11&12 \\\hline
1& 1.003 & 1.003 & 1.003 & 1.031 & 1.005 & 1.008 & 1.019 & 1.093 & 1.029 & 1.054 & 1.081 & 1.23 \\
2&1.003 & 1.003 & 1.028 & 1.235 & 1.005 & 1.198 & 1.252 & 1.437 & 1.05 & 1.196 & 1.246 & 1.395 \\
3& 1.002 & 1.024 & 1.52 & 1.751 & 1.009 & 1.25 & 1.753 & 1.833 & 1.049 & 1.214 & 1.643 & 1.753 \\
4&0.984 & 1.133 & 1.71 & \textbf{1.901} & 0.985 & 1.268 & 1.782 & 1.867 & 0.992 & 1.165 & 1.686 & 1.796 \\
5& 1.005 & 1.007 & 1.026 & 1.108 & 1.1 & 1.414 & 1.59 & 1.685 & 1.214 & 1.328 & 1.321 & 1.433 \\
6&1.006 & 1.197 & 1.285 & 1.442 & 1.379 & \cellcolor{gray!25}\textbf{1.714} & 1.792 & 1.819 & 1.291 & 1.614 & 1.733 & 1.764 \\
7& 1.002 & 1.218 & 1.753 & 1.834 & 1.448 & 1.683 & \textbf{1.818} & 1.862 & 1.225 & 1.627 & 1.769 & 1.835 \\
8&0.973 & 1.265 & 1.78 & 1.866 & 1.354 & 1.559 & 1.764 & 1.866 & 1.232 & 1.5 & 1.714 & 1.834 \\
9& 1.028 & 1.058 & 1.081 & 1.216 & 1.213 & 1.336 & 1.306 & 1.434 & 1.215 & 1.552 & 1.711 & 1.734 \\
10&1.045 & 1.196 & 1.247 & 1.392 & 1.286 & 1.613 & 1.733 & 1.756 & 1.51 &\textbf{1.685} & 1.762 & 1.806 \\
11&1.047 & 1.212 & 1.643 & 1.741 & 1.239 & 1.628 & 1.769 & 1.834 & 1.572 & 1.678 & \textbf{1.77} & 1.825 \\
12& 1.002 & 1.163 & 1.701 & 1.796 & 1.229 & 1.506 & 1.715 & 1.835 & 1.455 & 1.605 & 1.709 & 1.815 \\
\end{tabular}}\end{table}

\begin{table}[h]
\caption{Gain Matrix. Short horizon ($T=10000$)}%
\label{Tshort}%
\centering
\centering
\vspace{0.3cm} \scalebox{0.7}{
\begin{tabular}{c|cccccccccccc}
\text{$a_1\backslash a_2 $} & 1 & 2 & 3 & 4 & 5 & 6 & 7 & 8 &9&10&11&12 \\\hline
1&1.003 & 1.003 & 1.004 & 1.026 & 1.003 & 1.008 & 1.017 & 1.095 & 1.037 & 1.06 & 1.086 & 1.214 \\
2&1.003 & 1.003 & 1.004 & 1.026 & 1.008 & 1.124 & 1.215 & 1.347 & 1.073 & 1.192 & 1.252 & 1.386 \\
3&1.002 & 1.002 & 1.281 & 1.345 & 1.008 & 1.261 & 1.503 & 1.678 & 1.073 & 1.211 & 1.313 & 1.562 \\
4&0.981 & 0.981 & 1.316 &\textbf{1.901} & 0.975 & 1.242 & 1.613 & 1.864 & 1.027 & 1.17 & 1.394 & 1.737 \\
5&1.003 & 1.01 & 1.025 & 1.091 & 1.079 & 1.163 & 1.2 & 1.48 & 1.119 & 1.198 & 1.322 & 1.388 \\
6&1.006 & 1.122 & 1.298 & 1.394 & 1.148 &\textbf{1.563} & 1.641 & 1.705 & 1.179 & 1.513 & 1.621 & 1.723 \\
7& 0.999 & 1.189 & 1.503 & 1.678 & 1.141 & 1.553 &\cellcolor{gray!25}\textbf{1.774} & 1.806 & 1.198 & 1.548 & 1.739 & 1.792 \\
8& 0.97 & 1.195 & 1.6 & 1.863 & 1.234 & 1.475 & 1.713 & \textbf{1.859} & 1.202 & 1.474 & 1.69 & 1.832 \\
9&1.037 & 1.082 & 1.107 & 1.227 & 1.121 & 1.204 & 1.259 & 1.387 & 1.231 & 1.498 & 1.616 & 1.653 \\
10&1.051 & 1.192 & 1.241 & 1.373 & 1.174 & 1.512 & 1.635 & 1.717 & 1.459 &\textbf{1.658} & 1.738 & 1.78 \\
11&1.054 & 1.222 & 1.312 & 1.512 & 1.235 & 1.535 & 1.74 & 1.803 & 1.495 & 1.657 & \textbf{1.748} & 1.811 \\
12&0.994 & 1.161 & 1.461 & 1.733 & 1.182 & 1.48 & 1.676 & 1.831 & 1.398 & 1.584 & 1.698 & 1.81 \\
\end{tabular}}\end{table}

In both cases, all equilibria (indicated in bold) are conducive to
cooperation. We also indicate in gray the selection obtained by looking at the
(uniquely defined) limit QRE. For these matrices, the logit parameter can be
raised without bound, and we thus select an equilibrium. This gives us
$(\alpha^{*},b^{*})=(0.3,0.02)$ for the long horizon, and $(\alpha^{*}%
,b^{*})=(0.3,0.04)$ for the shorter one.

\subsubsection*{Stochastic Payoffs}

We report in Table~\ref{Tcor} (and \ref{Tind}) the gain matrices obtained when
payoffs are perfectly correlated (respectively independent), as we vary $y$,
setting $\alpha=0.1$ and $\epsilon=0.1$. \begin{table}[h]
\caption{Gain matrices, correlated payoffs}%
\label{Tcor}
\begin{subtable}{.33\linewidth}
\centering
\caption{$y=-0.8$}
\scalebox{0.55}{
\begin{tabular}{c|ccccccc}
$ \kappa _1\backslash \kappa _2$ & 0 & 1 & 2 & 3 & 4 &5&6\\\hline
0 & 1.11 & 1.14 & 1.24 & 1.31 & 1.46 & 1.56 & 1.67 \\
1 & 1.1 & 1.22 & 1.3 & 1.43 & 1.51 & 1.58 & 1.69 \\
2 & 1.14 & 1.25 & 1.39 & 1.51 & 1.63 & 1.67 & 1.75 \\
3 & 1.13 & 1.27 & 1.43 & 1.62 & 1.67 & 1.76 & 1.83 \\
4 & 1.12 & 1.23 & 1.44 & 1.58 & 1.75 & 1.82 & 1.85 \\
5 & 1. & 1.07 & 1.31 & 1.57 & 1.76 & 1.83 & 1.87 \\
6 & 0.71 & 0.85 & 1.06 & 1.48 & 1.74 & 1.82 &  \cellcolor{gray!25}1.87 \\
\end{tabular}
}
\end{subtable}\hfill\begin{subtable}{.33\linewidth}
\centering
\caption{$y=-0.6$}
\scalebox{0.55}{
\begin{tabular}{c|ccccccc}
$ \kappa _1\backslash \kappa _2$ & 0 & 1 & 2 & 3 & 4 &5&6\\\hline
0 & 1.1 & 1.18 & 1.2 & 1.34 & 1.46 & 1.54 & 1.77 \\
1 & 1.15 & 1.2 & 1.3 & 1.39 & 1.52 & 1.6 & 1.79 \\
2 & 1.1 & 1.23 & 1.39 & 1.49 & 1.61 & 1.71 & 1.8 \\
3 & 1.15 & 1.22 & 1.39 & 1.55 & 1.7 & 1.79 & 1.83 \\
4 & 1.09 & 1.21 & 1.37 & 1.6 & 1.75 & 1.82 & 1.87 \\
5 & 0.87 & 1.03 & 1.3 & 1.56 & 1.74 & 1.83 & 1.87 \\
6 & 0.61 & 0.79 & 1.05 & 1.43 & 1.71 & 1.84 & \cellcolor{gray!25} 1.88 \\
\end{tabular}
}
\end{subtable}\hfill\begin{subtable}{.33\linewidth}
\centering
\caption{$y=-0.4$}
\scalebox{0.55}{
\begin{tabular}{c|ccccccc}
$ \kappa _1\backslash \kappa _2$ & 0 & 1 & 2 & 3 & 4 &5&6\\\hline
0 & 1.15 & 1.19 & 1.3 & 1.38 & 1.49 & 1.66 & 1.88 \\
1 & 1.15 & 1.21 & 1.34 & 1.44 & 1.53 & 1.68 & 1.87 \\
2 & 1.19 & 1.27 & 1.41 & 1.52 & 1.6 & 1.73 & 1.87 \\
3 & 1.14 & 1.25 & 1.42 & 1.57 & 1.72 & 1.79 & 1.87 \\
4 & 1.06 & 1.16 & 1.33 & 1.61 & 1.75 & 1.82 & 1.89 \\
5 & 0.89 & 0.98 & 1.2 & 1.5 & 1.75 &  \cellcolor{gray!25}1.84 & 1.88 \\
6 & 0.49 & 0.7 & 0.95 & 1.33 & 1.68 & 1.83 & 1.88 \\
\end{tabular}
}
\end{subtable}
\par
\medskip
\par
\begin{subtable}{.33\linewidth}
\centering
\caption{$y=-0.2$}
\scalebox{0.55}{
\begin{tabular}{c|ccccccc}
$ \kappa _1\backslash \kappa _2$ & 0 & 1 & 2 & 3 & 4 &5&6\\\hline
0 & 1.18 & 1.22 & 1.3 & 1.46 & 1.59 & 1.74 & 2.04 \\
1 & 1.17 & 1.3 & 1.36 & 1.49 & 1.59 & 1.74 & 2.01 \\
2 & 1.17 & 1.27 & 1.42 & 1.52 & 1.67 & 1.76 & 1.96 \\
3 & 1.2 & 1.27 & 1.4 & 1.6 & 1.73 & 1.83 & 1.93 \\
4 & 1.07 & 1.16 & 1.38 & 1.59 &  \cellcolor{gray!25}1.76 & 1.84 & 1.9 \\
5 & 0.8 & 0.93 & 1.17 & 1.5 & 1.71 & 1.85 & 1.89 \\
6 & 0.48 & 0.64 & 0.92 & 1.32 & 1.64 & 1.83 & 1.89 \\
\end{tabular}
}
\end{subtable}\hfill\begin{subtable}{.33\linewidth}
\centering
\caption{$y=0$}
\scalebox{0.55}{
\begin{tabular}{c|ccccccc}
$ \kappa _1\backslash \kappa _2$ & 0 & 1 & 2 & 3 & 4 &5&6\\\hline
0 & 1.18 & 1.28 & 1.35 & 1.47 & 1.62 & 1.86 & 2.13 \\
1 & 1.23 & 1.35 & 1.38 & 1.53 & 1.68 & 1.86 & 2.1 \\
2 & 1.2 & 1.28 & 1.4 & 1.58 & 1.7 & 1.86 & 2.05 \\
3 & 1.16 & 1.28 & 1.44 & 1.59 & 1.77 & 1.87 & 1.99 \\
4 & 1.02 & 1.17 & 1.33 & 1.59 &  \cellcolor{gray!25}1.79 & 1.85 & 1.92 \\
5 & 0.82 & 0.9 & 1.11 & 1.47 & 1.71 & 1.84 & 1.91 \\
6 & 0.58 & 0.7 & 0.94 & 1.26 & 1.66 & 1.83 & 1.9 \\
\end{tabular}
}
\end{subtable}\hfill\begin{subtable}{.33\linewidth}
\centering
\caption{$y=0.2$}
\scalebox{0.55}{
\begin{tabular}{c|ccccccc}
$ \kappa _1\backslash \kappa _2$ & 0 & 1 & 2 & 3 & 4 &5&6\\\hline
0 & 1.28 & 1.29 & 1.42 & 1.53 & 1.75 & 1.99 & 2.21 \\
1 & 1.23 & 1.33 & 1.41 & 1.59 & 1.73 & 1.98 & 2.18 \\
2 & 1.24 & 1.31 &  \cellcolor{gray!25}1.5 & 1.6 & 1.75 & 1.94 & 2.11 \\
3 & 1.16 & 1.28 & 1.44 & 1.63 & 1.76 & 1.91 & 2.02 \\
4 & 1.04 & 1.14 & 1.29 & 1.56 & 1.75 & 1.88 & 1.95 \\
5 & 0.79 & 0.89 & 1.14 & 1.45 & 1.74 & 1.87 & 1.92 \\
6 & 0.67 & 0.78 & 0.98 & 1.32 & 1.67 & 1.83 & 1.91 \\
\end{tabular}
}
\end{subtable}
\par
\medskip
\par
\begin{subtable}{.33\linewidth}
\centering
\caption{$y=0.4$}
\scalebox{0.55}{
\begin{tabular}{c|ccccccc}
$ \kappa _1\backslash \kappa _2$ & 0 & 1 & 2 & 3 & 4 &5&6\\\hline
0 & 1.3 & 1.35 & 1.47 & 1.61 & 1.85 & 2.1 & 2.25 \\
1 & 1.29 & 1.38 & 1.49 & 1.63 & 1.83 & 2.08 & 2.23 \\
2 & 1.27 & 1.38 &  \cellcolor{gray!25}1.52 & 1.65 & 1.82 & 2.02 & 2.16 \\
3 & 1.22 & 1.27 & 1.45 & 1.68 & 1.8 & 1.94 & 2.07 \\
4 & 1.04 & 1.14 & 1.34 & 1.6 & 1.8 & 1.91 & 1.97 \\
5 & 0.87 & 0.97 & 1.18 & 1.47 & 1.74 & 1.88 & 1.93 \\
6 & 0.82 & 0.88 & 1.07 & 1.36 & 1.71 & 1.86 & 1.92 \\
\end{tabular}
}
\end{subtable}\hfill\begin{subtable}{.33\linewidth}
\centering
\caption{$y=0.6$}
\scalebox{0.55}{
\begin{tabular}{c|ccccccc}
$ \kappa _1\backslash \kappa _2$ & 0 & 1 & 2 & 3 & 4 &5&6\\\hline
0 & 1.35 & 1.43 & 1.54 & 1.73 & 1.99 & 2.18 & 2.29 \\
1 & 1.35 & \cellcolor{gray!25} 1.47 & 1.56 & 1.71 & 1.95 & 2.17 & 2.27 \\
2 & 1.29 & 1.4 & 1.55 & 1.69 & 1.9 & 2.09 & 2.21 \\
3 & 1.22 & 1.29 & 1.48 & 1.71 & 1.84 & 2. & 2.09 \\
4 & 1.06 & 1.17 & 1.35 & 1.62 & 1.81 & 1.93 & 2.01 \\
5 & 0.98 & 1.03 & 1.22 & 1.47 & 1.75 & 1.89 & 1.95 \\
6 & 0.94 & 1. & 1.15 & 1.43 & 1.72 & 1.87 & 1.93 \\
\end{tabular}
}
\end{subtable}\hfill\begin{subtable}{.33\linewidth}
\centering
\caption{$y=0.8$}
\scalebox{0.55}{
\begin{tabular}{c|ccccccc}
$ \kappa _1\backslash \kappa _2$ & 0 & 1 & 2 & 3 & 4 &5&6\\\hline
0 &  \cellcolor{gray!25}1.39 &  \cellcolor{gray!25}1.51 & 1.63 & 1.84 & 2.09 & 2.25 & 2.3 \\
1 &  \cellcolor{gray!25}1.43 &  \cellcolor{gray!25}1.48 & 1.6 & 1.83 & 2.05 & 2.21 & 2.28 \\
2 & 1.37 & 1.43 & 1.59 & 1.78 & 1.98 & 2.14 & 2.22 \\
3 & 1.28 & 1.37 & 1.51 & 1.72 & 1.9 & 2.04 & 2.11 \\
4 & 1.14 & 1.23 & 1.4 & 1.64 & 1.84 & 1.95 & 2.02 \\
5 & 1.09 & 1.17 & 1.3 & 1.56 & 1.78 & 1.91 & 1.98 \\
6 & 1.08 & 1.14 & 1.27 & 1.52 & 1.76 & 1.88 & 1.95 \\
\end{tabular}
}
\end{subtable}
\end{table}

\begin{table}[h]
\caption{Gain matrices, independent shocks}%
\label{Tind}
\begin{subtable}{.33\linewidth}
\centering
\caption{$y=-0.8$}
\scalebox{0.55}{
\begin{tabular}{c|ccccccc}
$ \kappa _1\backslash \kappa _2$ & 0 & 1 & 2 & 3 & 4 &5&6\\\hline
0 & 1. & 1.04 & 1.09 & 1.2 & 1.29 & 1.45 & 1.64 \\
1 & 1.01 & 1.06 & 1.14 & 1.23 & 1.37 & 1.5 & 1.67 \\
2 & 1. & 1.07 & 1.21 & 1.37 & 1.53 & 1.62 & 1.75 \\
3 & 1.02 & 1.08 & 1.28 & 1.53 & 1.72 & 1.77 & 1.85 \\
4 & 0.95 & 1.07 & 1.31 & 1.62 & 1.76 & \cellcolor{gray!15}1.84 & 1.87 \\
5 & 0.86 & 0.95 & 1.24 & 1.55 &\cellcolor{gray!15} 1.77 &\cellcolor{gray!25} 1.84 & 1.88 \\
6 & 0.57 & 0.77 & 1.06 & 1.5 & 1.74 & 1.84 & 1.87 \\
\end{tabular}
}
\end{subtable}\hfill\begin{subtable}{.33\linewidth}
\centering
\caption{$y=-0.6$}
\scalebox{0.55}{
\begin{tabular}{c|ccccccc}
$ \kappa _1\backslash \kappa _2$ & 0 & 1 & 2 & 3 & 4 &5&6\\\hline
0 & 1.04 & 1.06 & 1.15 & 1.21 & 1.34 & 1.5 & 1.76 \\
1 & 1.03 & 1.08 & 1.18 & 1.25 & 1.39 & 1.58 & 1.78 \\
2 & 1.05 & 1.1 & 1.25 & 1.42 & 1.54 & 1.69 & 1.82 \\
3 & 1. & 1.08 & 1.33 & 1.54 & 1.7 & 1.79 & 1.86 \\
4 & 0.93 & 1.04 & 1.28 & 1.57 &\cellcolor{gray!25} 1.76 & \cellcolor{gray!25}1.84 & 1.88 \\
5 & 0.8 & 0.95 & 1.21 & 1.52 & \cellcolor{gray!25}1.75 & \cellcolor{gray!25}1.85 & 1.88 \\
6 & 0.55 & 0.67 & 0.98 & 1.38 & 1.7 & 1.83 & 1.89 \\
\end{tabular}
}
\end{subtable}\hfill\begin{subtable}{.33\linewidth}
\centering
\caption{$y=-0.4$}
\scalebox{0.55}{
\begin{tabular}{c|ccccccc}
$ \kappa _1\backslash \kappa _2$ & 0 & 1 & 2 & 3 & 4 &5&6\\\hline
0 & 1.06 & 1.11 & 1.15 & 1.25 & 1.4 & 1.62 & 1.9 \\
1 & 1.07 & 1.08 & 1.19 & 1.37 & 1.49 & 1.64 & 1.88 \\
2 & 1.02 & 1.12 & 1.26 & 1.41 & 1.59 & 1.72 & 1.88 \\
3 & 1.01 & 1.18 & 1.29 & 1.49 & 1.68 & 1.82 & 1.91 \\
4 & 0.95 & 1.07 & 1.25 & 1.52 &\cellcolor{gray!25} 1.76 & 1.86 & 1.9 \\
5 & 0.75 & 0.89 & 1.15 & 1.52 & 1.73 & 1.86 & 1.89 \\
6 & 0.51 & 0.6 & 0.92 & 1.34 & 1.7 & 1.83 & 1.89 \\
\end{tabular}
}
\end{subtable}
\par
\medskip
\par
\begin{subtable}{.33\linewidth}
\centering
\caption{$y=-0.2$}
\scalebox{0.55}{
\begin{tabular}{c|ccccccc}
$ \kappa _1\backslash \kappa _2$ & 0 & 1 & 2 & 3 & 4 &5&6\\\hline
0 & 1.04 & 1.11 & 1.18 & 1.36 & 1.5 & 1.73 & 2.04 \\
1 & 1.06 & 1.13 & 1.23 & 1.37 & 1.53 & 1.75 & 2.02 \\
2 & 1.06 & 1.14 & 1.27 & 1.45 & 1.63 & 1.76 & 1.98 \\
3 & 1.07 & 1.12 & 1.33 & 1.56 & 1.73 & 1.85 & 1.94 \\
4 & 0.92 & 1.04 & 1.29 & 1.55 & 1.76 & 1.86 & 1.92 \\
5 & 0.77 & 0.85 & 1.1 & 1.46 & 1.72 & 1.85 & 1.91 \\
6 & 0.51 & 0.63 & 0.9 & 1.3 & 1.68 & 1.85 & 1.9 \\
\end{tabular}
}
\end{subtable}\hfill\begin{subtable}{.33\linewidth}
\centering
\caption{$y=0$}
\scalebox{0.55}{
\begin{tabular}{c|ccccccc}
$ \kappa _1\backslash \kappa _2$ & 0 & 1 & 2 & 3 & 4 &5&6\\\hline
0 & 1.09 & 1.17 & 1.25 & 1.38 & 1.61 & 1.88 & 2.14 \\
1 & 1.12 & 1.19 & 1.27 & 1.41 & 1.64 & 1.89 & 2.12 \\
2 & 1.09 & 1.17 & 1.35 & 1.48 & 1.68 & 1.88 & 2.07 \\
3 & 1.06 & 1.12 & 1.31 & 1.55 & 1.73 & 1.88 & 1.99 \\
4 & 0.96 & 1.03 & 1.25 & 1.51 & \cellcolor{gray!25}1.76 & 1.88 & 1.95 \\
5 & 0.78 & 0.83 & 1.09 & 1.45 & 1.73 & 1.86 & 1.92 \\
6 & 0.57 & 0.69 & 0.92 & 1.36 & 1.66 & 1.84 & 1.91 \\
\end{tabular}
}
\end{subtable}\hfill\begin{subtable}{.33\linewidth}
\centering
\caption{$y=0.2$}
\scalebox{0.55}{
\begin{tabular}{c|ccccccc}
$ \kappa _1\backslash \kappa _2$ & 0 & 1 & 2 & 3 & 4 &5&6\\\hline
0 & 1.16 & 1.2 & 1.31 & 1.51 & 1.73 & 2.01 & 2.22 \\
1 & 1.13 & 1.19 & 1.35 & 1.5 & 1.72 & 1.99 & 2.19 \\
2 & 1.12 & 1.22 & 1.32 & 1.56 & 1.76 & 1.97 & 2.14 \\
3 & 1.12 & 1.19 & 1.36 & \cellcolor{gray!25}1.57 & 1.77 & 1.93 & 2.04 \\
4 & 0.95 & 1.06 & 1.26 & 1.54 & 1.76 & 1.9 & 1.97 \\
5 & 0.78 & 0.89 & 1.1 & 1.44 & 1.73 & 1.87 & 1.93 \\
6 & 0.67 & 0.76 & 0.96 & 1.32 & 1.68 & 1.84 & 1.92 \\
\end{tabular}
}
\end{subtable}
\par
\medskip
\par
\begin{subtable}{.33\linewidth}
\centering
\caption{$y=0.4$}
\scalebox{0.55}{
\begin{tabular}{c|ccccccc}
$ \kappa _1\backslash \kappa _2$ & 0 & 1 & 2 & 3 & 4 &5&6\\\hline
0 & 1.22 & 1.27 & 1.41 & 1.59 & 1.85 & 2.14 & 2.28 \\
1 & 1.21 & \cellcolor{gray!25}1.3 & 1.41 & 1.59 & 1.84 & 2.11 & 2.23 \\
2 & 1.19 & 1.27 & 1.41 & 1.63 & 1.82 & 2.06 & 2.18 \\
3 & 1.13 & 1.19 & 1.41 &1.66 & 1.82 & 1.97 & 2.07 \\
4 & 0.99 & 1.08 & 1.26 & 1.53 & 1.8 & 1.92 & 1.98 \\
5 & 0.84 & 0.96 & 1.12 & 1.46 & 1.74 & 1.88 & 1.93 \\
6 & 0.79 & 0.89 & 1.07 & 1.38 & 1.71 & 1.86 & 1.93 \\
\end{tabular}
}
\end{subtable}\hfill\begin{subtable}{.33\linewidth}
\centering
\caption{$y=0.6$}
\scalebox{0.55}{
\begin{tabular}{c|ccccccc}
$ \kappa _1\backslash \kappa _2$ & 0 & 1 & 2 & 3 & 4 &5&6\\\hline
0 & \cellcolor{gray!25}1.28 & 1.34 & 1.5 & 1.72 & 1.99 & 2.21 & 2.29 \\
1 & 1.26 & 1.35 & 1.49 & 1.71 & 1.98 & 2.17 & 2.27 \\
2 & 1.23 & 1.32 & 1.47 & 1.71 & 1.94 & 2.11 & 2.2 \\
3 & 1.17 & 1.26 & 1.41 & 1.66 & 1.9 & 2.02 & 2.1 \\
4 & 1.05 & 1.12 & 1.3 & 1.58 & 1.81 & 1.94 & 2.01 \\
5 & 0.96 & 1.04 & 1.22 & 1.49 & 1.76 & 1.9 & 1.96 \\
6 & 0.93 & 1. & 1.16 & 1.44 & 1.71 & 1.88 & 1.94 \\
\end{tabular}
}
\end{subtable}\hfill\begin{subtable}{.33\linewidth}
\centering
\caption{$y=0.8$}
\scalebox{0.55}{
\begin{tabular}{c|ccccccc}
$ \kappa _1\backslash \kappa _2$ & 0 & 1 & 2 & 3 & 4 &5&6\\\hline
0 &\cellcolor{gray!25} 1.38 & 1.44 & 1.61 & 1.86 & 2.11 & 2.26 & 2.31 \\
1 & 1.36 & 1.42 & 1.59 & 1.83 & 2.08 & 2.23 & 2.28 \\
2 & 1.29 & 1.39 & 1.53 & 1.76 & 2.02 & 2.15 & 2.22 \\
3 & 1.21 & 1.32 & 1.5 & 1.7 & 1.9 & 2.06 & 2.13 \\
4 & 1.14 & 1.21 & 1.38 & 1.63 & 1.85 & 1.97 & 2.04 \\
5 & 1.1 & 1.16 & 1.31 & 1.55 & 1.78 & 1.91 & 1.98 \\
6 & 1.08 & 1.14 & 1.29 & 1.51 & 1.74 & 1.89 & 1.95 \\
\end{tabular}
}
\end{subtable}
\end{table}

\newpage

\subsubsection*{\label{app4}Duopoly}

Our duopoly example assumes $p=1.4+k0.1$ for $k\in\{0,...,6\}$. We reported in
the main text the payoff matrix for the stage game for $y=6$ and $10$. Tables
\ref{Tduo} reports expected gains when players use biased policy rules with
$b_{i}=\kappa_{i}\varpi$ where $\varpi=0.01$, for $\kappa_{i}\in\{0,...,6\}$.
Expected gains are obtained by taking the average of 8 simulations, each
running for 100000 periods, starting from unfavorable initial conditions.

\begin{table}[h]
\caption{Duopoly. Gain matrices}%
\label{Tduo}
\begin{subtable}{.5\linewidth}
\centering
\caption{$y=4$}
\scalebox{0.7}{
\begin{tabular}{c|ccccccc}
$ \kappa _1\backslash \kappa _2$ & 0 & 1 & 2 & 3 & 4 & 5&6 \\\hline
0 &\cellcolor{gray!25} 2.907 & \cellcolor{gray!25}2.994 & 3.095 & 3.283 & 3.373 & 3.437 & 3.545 \\
1 & \cellcolor{gray!25}2.932 & \cellcolor{gray!25}2.991 & 3.052 & 3.235 & 3.325 & 3.399 & 3.493 \\
2 & 2.698 & 2.878 & 2.998 & 3.109 & 3.234 & 3.322 & 3.42 \\
3 & 2.644 & 2.701 & 2.873 & 3.012 & 3.118 & 3.231 & 3.355 \\
4 & 2.61 & 2.662 & 2.777 & 2.918 & 3.06 & 3.147 & 3.282 \\
5 & 2.507 & 2.619 & 2.706 & 2.845 & 2.991 & 3.09 & 3.212 \\
6 & 2.4 & 2.541 & 2.651 & 2.78 & 2.901 & 3.021 & 3.146 \\
\end{tabular}
}
\end{subtable}\hfill\begin{subtable}{.5\linewidth}
\centering
\caption{$y=5$}
\scalebox{0.7}{
\begin{tabular}{c|ccccccc}
$ \kappa _1\backslash \kappa _2$ & 0 & 1 & 2 & 3 & 4 & 5 & 6\\\hline
0 & 2.588 & 2.941 & 3.019 & 3.257 & 3.437 & 3.539 & 3.657 \\
1 & 2.886 &\cellcolor{gray!25} \textbf{2.945} & 3.04 & 3.204 & 3.434 & 3.522 & 3.627 \\
2 & 2.748 & 2.924 & 2.994 & 3.098 & 3.337 & 3.465 & 3.572 \\
3 & 2.55 & 2.634 & 2.877 & 3.014 & 3.165 & 3.336 & 3.48 \\
4 & 2.534 & 2.588 & 2.71 & 2.913 & 3.043 & 3.189 & 3.361 \\
5 & 2.474 & 2.548 & 2.634 & 2.781 & 2.948 & 3.111 & 3.283 \\
6 & 2.298 & 2.477 & 2.582 & 2.697 & 2.844 & 2.999 & 3.171 \\
\end{tabular}
}
\end{subtable}
\par
\medskip\begin{subtable}{.5\linewidth}
\centering
\caption{$y=6$}
\scalebox{0.7}{
\begin{tabular}{c|cccccc}
$ \kappa _1\backslash \kappa _2$ & 0 & 1 & 2 & 3 & 4 & 5 \\\hline
0 &  2.233 & 2.436 & 2.939 & 3.248 & 3.553 & 3.778  \\
1 & 2.362 & \cellcolor{gray!25} \textbf{2.955} & 2.984 & 3.239 & 3.536 & 3.767 \\
2 & 2.673 & 2.851 & 2.944 & 3.142 & 3.423 & 3.685\\
3 & 2.42 & 2.576 & 2.876 & 3.051 & 3.24 & 3.536 \\
4 &  2.435 & 2.489 & 2.638 & 2.929 & 3.105 & 3.344 \\
5 & 2.366 & 2.469 & 2.544 & 2.73 & 2.941 & 3.193 \\
\end{tabular}
}
\end{subtable}\hfill\begin{subtable}{.5\linewidth}
\centering
\caption{$y=7$}
\scalebox{0.7}{
\begin{tabular}{c|ccccccc}
$ \kappa _1\backslash \kappa _2$ & 0 & 1 & 2 & 3 & 4 & 5 & 6\\\hline
0 & 2.119 & 2.14 & 2.525 & 2.937 & 3.357 & 3.671 & 3.891 \\
1 & 2.114 & 2.46 & 2.826 & 3.047 & 3.298 & 3.641 & 3.897 \\
2 & 2.363 & 2.738 & \cellcolor{gray!25}2.955 &\cellcolor{gray!25} 3.081 & 3.254 & 3.576 & 3.842 \\
3 & 2.488 & 2.748 & \cellcolor{gray!25}2.957 & \cellcolor{gray!25}3.031 & 3.152 & 3.431 & 3.713 \\
4 & 2.361 & 2.426 & 2.659 & 2.943 & 3.098 & 3.271 & 3.521 \\
5 & 2.362 & 2.376 & 2.465 & 2.719 & 2.995 & 3.164 & 3.371 \\
6 & 2.288 & 2.38 & 2.425 & 2.544 & 2.799 & 3.006 & 3.223 \\
\end{tabular}
}
\end{subtable}
\medskip
\par
\begin{subtable}{.5\linewidth}
\centering
\caption{$y=8$}
\scalebox{0.7}{
\begin{tabular}{c|ccccccc}
$ \kappa _1\backslash \kappa _2$ & 0 & 1 & 2 & 3 & 4 & 5&6 \\\hline
0 & 2.035 & 2.05 & 2.169 & 2.479 & 3.118 & 3.631 & 3.935 \\
1 & 2.034 & 2.151 & 2.504 & 2.703 & 3.184 & 3.574 & 3.912 \\
2 & 2.085 & 2.428 & 2.732 & 2.946 & 3.177 & 3.499 & 3.884 \\
3 & 2.16 & 2.458 & 2.839 & \cellcolor{gray!25}\textbf{3.043} & 3.119 & 3.399 & 3.763 \\
4 & 2.278 & 2.479 & 2.777 & 2.961 & 3.055 & 3.269 & 3.57 \\
5 & 2.251 & 2.291 & 2.426 & 2.718 & 2.988 & 3.168 & 3.404 \\
6 & 2.214 & 2.283 & 2.341 & 2.489 & 2.79 & 3.035 & 3.269 \\
\end{tabular}
}
\end{subtable}\hfill\begin{subtable}{.5\linewidth}
\centering
\caption{$y=9$}
\scalebox{0.7}{
\begin{tabular}{c|ccccccc}
$ \kappa _1\backslash \kappa _2$ & 0 & 1 & 2 & 3 & 4 & 5 & 6\\\hline
0 & 2.023 & 2.034 & 2.06 & 2.3 & 2.695 & 3.566 & 3.968 \\
1 & 2.024 & 2.072 & 2.244 & 2.428 & 2.796 & 3.5 & 3.928 \\
2 &2.017 & 2.189 & 2.427 & 2.793 & 3.117 & 3.462 & 3.842 \\
3 & 2.043 & 2.212 & 2.688 & \cellcolor{gray!25} \textbf{3.038} & 3.103 & 3.339 & 3.729 \\
4 & 2.061 & 2.262 & 2.82 & 2.983 & 3.061 & 3.23 & 3.554 \\
5 &  2.232 & 2.306 & 2.523 & 2.791 & 3.014 & 3.153 & 3.396 \\
6 &  2.161 & 2.226 & 2.293 & 2.466 & 2.761 & 3.034 & 3.293 \\
\end{tabular}
}
\end{subtable}
\par
\medskip
\par
\begin{subtable}{.5\linewidth}
\centering
\caption{$y=10$}
\scalebox{0.7}{
\begin{tabular}{c|ccccccc}
$ \kappa _1\backslash \kappa _2$ & 0 & 1 & 2 & 3 & 4 & 5&6 \\\hline
0 & 2.023 & 2.026 & 2.045 & 2.193 & 2.487 & 3.106 & 3.913 \\
1 & 2.019 & 2.028 & 2.066 & 2.308 & 2.563 & 3.229 & 3.861 \\
2 & 2.016 & 2.041 & 2.251 & 2.721 & 2.992 & 3.284 & 3.787 \\
3 & 2.025 & 2.127 & 2.622 & \cellcolor{gray!25}2.803 &\cellcolor{gray!25} 3.032 & \cellcolor{gray!15}3.26 & 3.65 \\
4 & 1.959 & 2.116 & 2.726 & \cellcolor{gray!25}2.922 & \cellcolor{gray!25}3.01 & \cellcolor{gray!15}3.193 & 3.519 \\
5 & 2. & 2.246 & 2.538 &\cellcolor{gray!15} 2.841 & \cellcolor{gray!15}3.012 & \cellcolor{gray!10}3.119 & 3.402 \\
6 & 2.097 & 2.174 & 2.29 & 2.551 & 2.805 & 3.029 & 3.285 \\
\end{tabular}
}
\end{subtable}\hfill\begin{subtable}{.5\linewidth}
\centering
\caption{$y=11$}
\scalebox{0.7}{
\begin{tabular}{c|ccccccc}
$ \kappa _1\backslash \kappa _2$ & 0 & 1 & 2 & 3 & 4 & 5 & 6\\\hline
0 & 2.018 & 2.025 & 2.041 & 2.112 & 2.392 & 2.809 & 3.816 \\
1 & 2.019 & 2.021 & 2.052 & 2.181 & 2.446 & 2.867 & 3.753 \\
2 & 2.017 & 2.032 & 2.165 & 2.483 & 2.701 & 2.99 & 3.701 \\
3 & 2.022 & 2.041 & 2.389 & \cellcolor{gray!25}2.829 & \cellcolor{gray!25}2.961 & \cellcolor{gray!25}3.159 & 3.596 \\
4 & 1.939 & 2.05 & 2.426 & \cellcolor{gray!25}2.843 &\cellcolor{gray!25} 2.965 & \cellcolor{gray!25}3.137 & 3.477 \\
5 & 1.843 & 2.026 & 2.347 &\cellcolor{gray!25} 2.844 & \cellcolor{gray!25}3.002 & \cellcolor{gray!25}3.121 & 3.355 \\
6 & 2.076 & 2.163 & 2.349 & 2.617 & 2.905 & 3.043 & 3.271 \\
\end{tabular}
}
\end{subtable}
\end{table}

\end{document}